\documentclass[12pt,preprint]{aastex}
\newcommand{\beq}{\begin{equation}}
\newcommand{\eeq}{\end{equation}}
\def\div{\vec\nabla\cdot}
\def\grad{\vec\nabla}

\def\rar{\rightarrow}

\def\az{a_0}
\def\msun{{\rm M}_{\odot}}

\def\cmst{~{\rm cm~ s}^{-2}~}
\def\gcmt{~{\rm g~ cm}^{-2}~}

\def\Sz{\Sigma_0}
\def\S{\Sigma}
\def\a{\alpha}

\def\kpc{{\rm kpc}}
\def\mpc{{\rm Mpc}}
\def\gpc{{\rm Gpc}}

\def\vg{{\bf g}}

\def\vr{{\bf r}}

\begin{document}

\title{Rings and shells of ``dark matter''  as MOND artifacts}
\author{Mordehai Milgrom\altaffilmark{1} and Robert H. Sanders\altaffilmark{2}}
\altaffiltext{1}{Center for Astrophysics, Weizmann Institute of
Science Rehovot 76100, Israel} \altaffiltext{2}{Kapteyn Astronomical
Institute, 9700 AV Groningen, Netherlands}

\begin{abstract}

MOND predicts that a mass, $M$, contained within its transition
radius $r_t\equiv (MG/\az)^{1/2}$, may exhibit a feature at about
that radius in the form of a shell, or projected ring, in the
deduced distribution of its phantom dark matter. This occurs despite
the absence of any underlying feature in the true (``baryon'')
source distribution itself. The phenomenon is similar to the
appearance of an event horizon and other unusual physics ``in the
middle of nothing'' near the transition radius of General Relativity
$MG/c^2$. We consider the possibility that this pure MOND phenomenon
is in the basis of the recent finding of such a ring in the galaxy
cluster Cl 0024+17 by Jee et al. We find that the parameters of the
observed ring can be naturally explained in this way; this feature
may therefore turn out to be a direct evidence for MOND. We study
this phenomenon in simple, axisymmetric configurations aligned with
the line of sight: spherical masses, a dumbbell of spherical masses,
and an elongated, thin structure. The properties of the apparent
ring: its radius, surface density, and contrast, depend on the form
of the MOND interpolating function and on the exact three
dimensional distribution of the sources (the thin-lens approximation
is quite invalid in MOND). We also comment on the possible
appearance of orphan features, marking the Newtonian-to-MOND
transition, in high surface brightness galaxies. In particular, we
find that previously unexplained structure in the rotation curves of
some galaxies may be evidence for such features.

\end{abstract}


\section{Introduction}

 Prompted by the recent claim of a ring-like feature in the
surface density distribution of the galaxy cluster Cl 0024+17, as
deduced from weak-lensing analysis by Jee et al. (2007), we
investigate the occurrence of such features in the modified
Newtonian dynamics (MOND; Milgrom 1983a; for reviews see Sanders and
McGaugh 2002, Bekenstein 2006, Milgrom 2008). In MOND, unlike
Newtonian dynamics, a mass $M$ defines a natural scale length
$r_t\equiv (MG/\az)^{1/2}$ that is unrelated to a scale-length in
the mass distribution itself ($\az$ is the acceleration constant
introduced by MOND). This is the so called transition radius
(Milgrom 1983a,b, 1986a). For a concentrated mass it marks the
transition from the Newtonian regime at small radii to the MOND
regime at large radii. When the mass $M$ is much more extended than
its transition radius, no special effects are expected at $r_t$. In
this case the acceleration inside the body is smaller than $\az$ and
the system is everywhere in the deep MOND regime: there is no
Newtonian-MOND transition anywhere. However, when the mass is well
within its transition radius there is a marked change in the
dynamics of test particles when crossing the transition region: the
potential goes from $1/r$ to logarithmic; the rotation curve goes
from Keplerian to flat, etc.. If we then look at the fictitious
``dark matter'' density distribution, or, as is done with lensing,
at its projected surface density distribution, there may appear near
this radius a pronounced feature in the form of a maximum. For an
axisymmetric system aligned with the line of sight this shows as a
``ring'' of dark matter. All this occurs without there being any
corresponding feature in the source (baryonic) distribution.

Physics is replete with similar examples in which a feature appears
in some field quantity at a radius that is unrelated to a
length-scale characterizing the  source distribution of the field.
As in our case, the radius of the feature depends on the strength of
the source alone and on some constants appearing in the field
equations. One example is the appearance of a horizon at the
characteristic scale attached to a mass in relativity, which is
introduced by $c$ and $M$; i.e., the gravitational radius $MG/c^2$.
The Bohr radius, $r_B$, as a mark of a transition from quantum
behavior for $r< r_B$ to a classical one for $r\gg r_B$ is another.
And in less fundamental examples: the Bondi radius in spherical
accretion, which depends on the central mass and the ambient gas
temperature; and the screening length for a charge $Q$ in a plasma.

In this regard it is interesting to note that, because of the
approximate equality $\az\sim cH_0$ (Milgrom
1983a), we can write
 \beq r_t\sim (R_sR_H)^{1/2}, \label{ia} \eeq
  where
$R_s$ is the Schwarzschild radius of the mass, and $R_H=c/H_0$ is
the Hubble radius.  The transition radius is thus, approximately,
the harmonic mean of these two horizon radii. It is then of the
order of the Einstein radius for cosmological lenses, but is much
larger for local ones.
\par
Here we consider this interesting MOND phenomenon in more detail. It
turns out that beside the obvious dependence on the mass
distribution in the lens, the appearance of the above feature
depends sensitively on the behavior of the yet undetermined
interpolating function of MOND around its transition region. We thus
calculate the ring for different forms of this function.
\par
Regarding the implications of their finding to MOND, Jee et al
(2007) make the following statement: ``The ringlike mass structure
at r = 0.4 Mpc surrounding the dense core at r$\le  0.25$ Mpc not
traced by the cluster ICM nor by the cluster galaxies serves as the
most definitive evidence from gravitational lensing to date for the
existence of dark matter. If there is no dark matter and the cluster
ICM is the dominant source of gravity, the MONDian gravitational
lensing mass should follow the ICM, which, however, does not show
any hints of such peculiar mass distribution.'' Although many were
quick to embrace this view,  it is, in fact, quite baseless: such
orphan rings of phantom dark matter (PDM) are formed naturally in
MOND (as possibly in other modified dynamics theories). Although
they tend to be overwhelmed by the underlying baryon distribution
itself (as they would in the DM scenario proper), there are
circumstances in which they can be observed directly (i.e., without
subtracting the source distribution).

As regards the ring in Cl 0024+17, we find that it can  be
reproduced naturally with this MOND phenomenon.  It will be exciting
indeed (and ironical) if the ring  discovered by Jee et al. turns
out to be a direct image of the MOND transition region, akin to an
image of an event horizon of a black hole as viewed via its lensing
effect on background sources.

This is an opportunity to dispel a common misconception about MOND:
it does not predict that the PDM distribution follows that of the
(baryon) sources as stated, e.g., in the above quotation from Jee et
al. and in the discussion of the implications of the bullet cluster
for MOND (Clowe et al. 2006, and see Angus et al. 2007 for the MOND
answer). An obvious counterexample is a baryonic thin disc, which is
predicted to have both a PDM disc of finite mass (whose mass
distribution does not follow that of the source disc) and an
extended spheroid of PDM which clearly doesn't follow the source
distribution (Milgrom 2001). Another example is a simple system of
several point masses, whose distribution of MOND PDM is very
complex, with, among other things, numerous regions of negative PDM
density (Milgrom 1986b) and a system of surfaces of maximum density
(such as the shells we discuss here).

Jee et al. (2007) attribute the observed ring to an actual ring (in
3-D) of dark matter resulting from a collision of two sub-clusters
along the line of sight. Famaey et al. (2007a) note that MOND has
long been known to require that the cores of clusters be dominated
by some form of as yet undetected matter (e.g., Gerbal et al. 1992,
Sanders 1999, Aguirre, Schaye \& Quataert 2001, Sanders 2003,
Pointecouteau \& Silk 2005, Angus Famaey \& Buote 2007). Famaey et
al. (2007a) considered, in particular, neutrinos as proposed by
Sanders (2003). They thus show that the explanation of Jee et al.
can be adopted in the MOND framework without adding new ingredients.
Angus et al. (2007) show that this as yet unseen matter also
accounts for the observations of the bullet cluster (Clowe et al.
2006).

The explanation of the observed ring as a projection of the MOND PDM
transition shell does not require, of course, a collision to have
produced it. However, it is possible that such a collision has
created a mass distribution that brings out the ring more clearly.
We also find that, although it is not necessary, an aligned dumbbell
configuration is more conducive to the detectability of a ring than
a single spherical mass.

Our study here is by no means  exhaustive; we only aim at
demonstrating the phenomenon and studying in broad terms how it
depends on the various inputs. In particular, we limit ourselves
to some simple axisymmetric mass distribution that, in addition,
are aligned with the line of sight. This applies to spherical
systems in general, and also to the case of the cluster Cl
0024+17, which is believed to be an aligned double cluster.

In section 2 we lay out the basic ideas behind the appearance of
``rings'' in MOND. In section 3 we give the results for various
axisymmetric systems and various forms of the interpolating
function. In section 4 we consider the possible relevance to Cl
0024+17. Section 5 discusses the phenomenon in single galaxies.
Section 6 is a discussion.

\section{The phantom ``dark matter'' surface density: shells and rings}

When studying the potential field of a mass distribution in MOND, it
is sometime useful to describe it in terms of the added mass that
would be required to produce the same field in Newtonian dynamics.
This will remain useful as long as many continue to think in terms
of dark matter. In the present nonrelativistic formulations of MOND
as modified gravity,  this phantom dark matter (PDM) density,
$\rho_p$, is given by (Milgrom 1986b, 2001)
 \beq \rho_p=-(4\pi G)^{-1}\div\vg-\rho.\label{i}\eeq
Here $\rho$ is the actual (``baryon'') mass density, and
$\vg(\vr)$ is the MOND acceleration field. The first term in
eq.(\ref{i}) is the dynamical mass density, $\rho_D$, deduced
using Newtonian dynamics.

Properties of the PDM distribution have been investigated in the
past. For example, Milgrom (1986b) showed that the PDM density can
be negative under certain circumstances; Brada and Milgrom (1999)
showed that it cannot produce accelerations much exceeding $\az$ no
matter what the true density distribution is, and this was confirmed
by Milgrom and Sanders (2005) for a sample of galaxies from rotation
curve analysis; Milgrom (2001) studied the properties of the PDM
halo of disc galaxies. Here we shall concentrate on another property
of the PDM halo and show that its density can have maximal surfaces
in regions of space where no such features exist in the underlying
source distribution $\rho$ (they may, e.g., appear in vacuum). And
when projected on the sky these surfaces lead to the appearance of
line features such as rings or other ridge-like structures.

The total Newtonian dynamical mass of an isolated (bounded) mass is infinite
 as the MOND potential is asymptotically logarithmic in
this case. However, in some sense we can say that two density
distributions that have the same total (bounded) mass, also have the
same total dynamical mass: Consider two density distributions
$\rho_1$ and $\rho_2$ having the same total (true) mass, and both
bound. Take a large volume $\ell U$ of a fixed shape $U$ but
increasing linear dimension $\ell$. Then the resulting dynamical (or
phantom) mass distributions satisfy in the limit $\ell\rar \infty$
 \beq \int_{\ell U}(\rho_{D,1}-\rho_{D,2})d^3r\rar 0.
 \label{xxi}\eeq

 This is because from eq.(\ref{i})
 \beq \int_{\ell U}(\rho_{D,1}-\rho_{D,2})d^3r\propto\int_{\ell S}(\vg_1-\vg_2)\cdot d\vec \sigma,
 \label{xxii}\eeq
where $S$ is the surface of $U$. And, Milgrom (1986a) has shown that
next to the leading term at large radii, $-(MG\az)^{1/2}\vec r/r^2$,
which is common to $\vg_1$ and $\vg_2$, there is a term (which may
be angle dependent) that decreases as $r^{-(\sqrt{3}+1)}$, fast
enough to make the limit of expression (\ref{xxii}) vanish.

In the Lagrangian formulation, put forth by Bekenstein and Milgrom
(1984), we have $\vg=-\grad\phi$, where $\phi$ is the gravitational
potential satisfying the modified, nonlinear Poisson equation
 \beq \div[\mu(|\grad\phi|/\az|)\grad\phi]=4\pi G\rho, \label{pois}\eeq
where $\mu(x)$ is the MOND interpolating function satisfying
$\mu(x)\rar 1$ for $x\rar\infty$, and $\mu(x)\approx x$ for
$x\rar0$.  We shall, however use throughout the much more manageable
approximation to this theory in which the MOND and Newtonian
accelerations, $\vg$ and $\vg_N$ respectively, are algebraically
related:
 \beq  \mu(|\vg |/\az)\vg=\vg_N.\label{ii}\eeq
It is sometimes more convenient to work with the inverse of
eq.(\ref{ii}) and define the interpolating function $\nu(y)$
(Milgrom 1986a) such that
 \beq  \vg=\nu(|\vg_N |/\az)\vg_N.\label{iia}\eeq

 The algebraic relation was the first formulation of
the MOND paradigm, posited to apply for test particle motion
(Milgrom 1983a). It is an exact consequence of the Lagrangian
formulation, for example, for spherical systems and it captures the
salient consequences of the Lagrangian formulation in other
configurations. At any rate it is good enough for our demonstrative
purposes here.
 Note that $\mu(x)=1/\nu(y)$, where $x=g/\az$,
and $y=g_N/\az$, and that $\nu(y)\rar 1$ for $y\rar \infty$ and
$\nu(y)\rar y^{-1/2}$ for $y\rar 0$.

The phenomenon discussed here may be studied with massive test
particles, as in rotation-curve analysis. But, since the present
motivation comes from lensing, we discuss the phenomenon in terms of
its appearance in lensing analysis: we present the results in the
form of the projected surface density derived from $\rho_p$, or of
that of the total dynamical mass density $\rho_D$.

In calculating lensing in MOND we assume that it is given by the
standard general relativistic formula using the non-relativistic
MOND potential, which governs the motion of massive test particles.
This recipe follows from TeVeS (Bekenstein 2004), the best
relativistic formulation of MOND we have at present.

As in Milgrom (1986a), we exploit the scaling relations of the MOND
equations (both formulations enjoy the same relations) and use
henceforth dimensionless quantities: Our unit of mass will be $M$,
the total true (``baryonic'') mass of the system; the unit of length
is the transition radius, $r_t\equiv (MG/\az)^{1/2}$. All other
units are formed from them: that of acceleration is $MG/r^2_t=\az$;
that of the gravitational potential is  $\az
r_t=V^2_{\infty}=(MG\az)^{1/2}$; that of density is $Mr^{-3}_t$; and
that of surface density is $\Sz\equiv M r^{-2}_t=\az/G$. In these
units $G$ and $\az$ disappear from the above MOND equations. Note
that unlike the units of other quantities, which depend on the total
mass, those for acceleration and surface density are universal
constants. This fact underlies the results of Brada and Milgrom
(1999) and also our results here showing that $\Sz$ is always the
characteristic surface density of the PDM in the region of the
feature. Note also that because of the nearness of $\az$ to
cosmological acceleration (Milgrom 1983a) $\Sz$ is also on the order
of the critical lensing surface density for objects at cosmological
distance (Sanders 1999).

It is easy to understand why $\rho_p$, and hence possibly also
$\rho_D$, may possess a shell-like feature: This occurs when the
mass in the system is well contained within its transition radius.
Let us consider, heuristically, a point mass, and assume a MOND
interpolating function of the form that crosses abruptly from one
asymptotic regime to the other:
\beq\mu(x)=\left\{\begin{array}{l@{\quad:\quad} l}1 & x\ge 1\\x & x<
1\end{array}\right. .\label{iii}\eeq
 Then,  for all radii smaller than 1
the accelerations are in the Newtonian regime, and no PDM is
predicted there. Beyond $r=1$ we have already deep MOND behavior;
so, $\rho_p$ jumps to $1/4\pi$ at $r=1$ and hence outward
$\rho_p=1/4\pi r^2$: a shell in the PDM distribution. The
corresponding projected surface density that will be deduced from
standard GR lensing in this case is:
 \beq \Sigma(R)={\pi \over
2}\Sigma(0)R^{-1}\left\{\begin{array}{l@{\quad:\quad} l}1-{2\over
\pi}arctan[(R^{-2}-1)^{1/2}] & R\le 1\\1 & R>1\end{array}\right.
,\label{iv}\eeq

where the central value $\S(0)=1/2\pi$, and $R$ is the projected
radius.

Thus $\S(R)$ (for the PDM alone) starts at $1/2\pi$ at the center,
increases to $1/4$  at $R=1$ and beyond it declines as $1/4R$. This
is the ``ring'' we speak of. This example already demonstrates the
salient properties of the ring in the more general (axisymmetric)
cases: The ring appears at a radius of order 1, has a maximum
surface density of order (usually somewhat smaller than) 1, and has
a contrast of that order.

Because the appearance of the shell feature is contingent on the
transition from the Newtonian regime to the MOND regime, its exact
properties depend strongly on the behavior of the transition
function $\mu(x)$ near $x=1$. This is also true of the exact maximum
acceleration the PDM can create as discussed in Brada and Milgrom
(1999) (the two phenomena are related). If such MOND rings are
clearly identified in enough systems this may provide important
constraints on the form of $\mu$ in the transition region. As part
of our study we consider various forms of $\mu$. We next discuss
this function in more detail.

\subsection{The MOND interpolating function}
In modified inertia formulations of MOND it is not clear that there
is one transition function that characterizes all motions, although
such a function does appear universally in the description of
circular motions in axisymemtric potentials (Milgrom 1994). From
this it would follow that a shell in the PDM distribution as deduced
from rotation curve analysis, but not necessarily lensing analyses,
is expected also in modified inertia formulations of MOND.  In fact,
given the sensitivity of the appearance of the ring to the form of
the interpolating function, the ring may well have a different
location or amplitude in lensing vs. rotation curve analysis in the
context of modified intertia formulations (this could possibly
distinguish between modified gravity and modified inertia). However,
such formulations are not yet developed enough to tell us what to
expect when probing the potential field with non circular motions
(such as lensing).
 So, here we base our discussion on modified gravity
formulations, in which, generically, an interpolating function
appears. For example, in the above Lgrangian theory, the
Lagrangian density of the gravitational potential is $\propto
F[(\grad\phi)^2/\az^2]$, and $\mu(x)\propto dF(x^2)/d(x^2)$.

We consider representatives from three one-parameter families of
interpolating functions: The first is
 \beq \mu_\a(x)={x\over (1+x^\a)^{1/\a}}, \label{v}\eeq
with the corresponding
 \beq \nu_\a(y)=\left[{1+(1+4y^{-\a})^{1/2}\over 2}   \right]^{1/\a}. \label{va}\eeq
 The case $\a=2$ has been extensively used in
rotation curve analysis from the first analysis by Kent (1987),
through Begeman Broeils and Sanders (1991), and to this day. The
choice $\mu_1$ has been used occasionally in various analyses (e.g.
Milgrom 1986a) and has recently been suggested to perform better
than $\mu_2$ in rotation curve analysis (Zhao and Famaey 2006,
Sanders and Noordermeer 2007). Famaey and Binney (2005) found that
neither $\mu_1$ nor $\mu_2$ adequately fit the rotation curve of the
Milky Way with its Basel mass model (Bissantz, Englmaier and Gerhard
2003); a good fit is achieved for a function that is near $\mu_1$
for small arguments but approaches $\mu_2$ at high arguments. The
limiting case, eq.(\ref{iii}), corresponds to very large $\a$.
Moreover, where $y\rar\infty,$ $\nu_{\a}\approx 1+y^{-\a}/\a$; for
$y\rar 0,$ $\nu_{\a}\approx y^{-1/2}+y^{(\a-1)/2}/2\a.$

The second and third families, which we consider here for
the first time, are defined by their inverse, or $\nu(y)$, functions:
 \beq \tilde\nu_{\a}(y)=(1-e^{-y})^{-1/2}+\a e^{-y}, \label{vi}\eeq
 with its corresponding $\tilde\mu_{\a}(x)$,
and
 \beq
 \bar\nu_{\a}(y)\equiv(1-e^{-y^\a})^{(-1/2\a)}+(1-1/2\a)e^{-y^\a},\label{vii}\eeq
 with the corresponding $\bar\mu_{\a}$.
At small $y$ this third family goes as
$y^{-1/2}+y^{\a-1/2}/4\a+1-1/2\a$. The choice of $\a$ permits a very
fast transition to the Newtonian regime near $y=1$.

These choices do not follow from any physical considerations. [In
fact, at present, we know of no concrete theoretical considerations
that can dictate the form of $\mu(x)$.] They do not even span the
whole range of possibilities. We only use them to show the variety
that is possible and to demonstrate the sensitivity of the ring
phenomenon to the exact choice.

There are some constraints on the interpolating function: $\mu\rar
1$ for $x\rar\infty$ is dictated by Newtonian correspondence
(although the way in which mu approaches 1 is not known, but enters
crucially into phenomena such as solar system effects). The limit
$\mu\rar x$ for $x\rar0$ is dictated by the basic premises of MOND
(it insures asymptotically flat rotation curves for isolated
systems). Another requirement is that $\hat\mu(x)\equiv
dln(\mu)/dln(x)>-1$ everywhere. This is the ellipticity condition
for the Lagrangian theory field equation. It is tantamount to the
sign definiteness of the Hessian of the Lagrangian as a function of
the components of $\grad\phi$) and insures uniqueness of the
solution under the Dirichlet and/or Neumann boundary condition (e.g.
Milgrom 1986a). This condition also says that $x\mu(x)$ [and
equivalently $y\nu(y)$] is a monotonic function, which is required
for the above algebraic relation, eqs.(\ref{ii})(\ref{iia}), between
$\vg$ and $\vg_N$ to have a unique solution. Beyond this requirement
the form of $\mu$ is free at the moment and we can only hope to use
the data to constrain it.

Zhao and Famaey (2006) have pointed out that in the formulation of
TeVeS in Bekenstein (2004) the effective interpolating function
$\mu_e(x)$ has to approach 1 rather slowly in the high acceleration
limit. Their argument is best paraphrased in terms of the $\nu$
function: If $\nu$ is that of the scalar field in TeVeS, then
$\nu_e(y)=\nu(y)+1$ is the effective function for the full gravity.
At small $y$, $\nu\approx y^{-1/2}$ as dictated by MOND, so
$y\nu(y)$ is increasing there. But the ellipticity condition says
that $y\nu(y)$ has to be a monotonic function; so it must remain
increasing, and hence $\nu$ cannot vanish faster than $a/y$ at large
$y$. So, in turn, $\nu_e$ cannot approach 1 faster than that. Since
$\mu(x)=1/\nu(y)$, and for $y\gg1$ $x\approx y$, it follows that
also $\mu(x)$ has to approach 1 no faster than $a/x$ (see also
Famaey et al. 2007b for a discussion of this constraint and for ways
to relax it). None of the interpolating functions considered above
satisfy this requirement except for $\mu_1$.  We, however, ignore
this kind of constraint. It is exactly this behavior that
potentially gets TeVeS in trouble with solar system and binary
pulsar constraints (see e.g. the detailed discussion in Bruneton and
Esposito-Farese 2007). So, a working version of TeVeS may well have
to have this aspect of it expurgated. It may be possible to dispose
of this constraint altogether by considering a theory with other
potentials. For example the relativistic version of Zlosnik Ferreira
\& Starkman 2007 doesn't seem to require this behavior. From a
theoretical point of view, there remains complete freedom to choose
the form of $\mu$ so long as the basic constraints outlined above or
satisfied.

\begin{figure}
\begin{tabular}{rl}
\tabularnewline
\includegraphics[width=0.5\columnwidth]{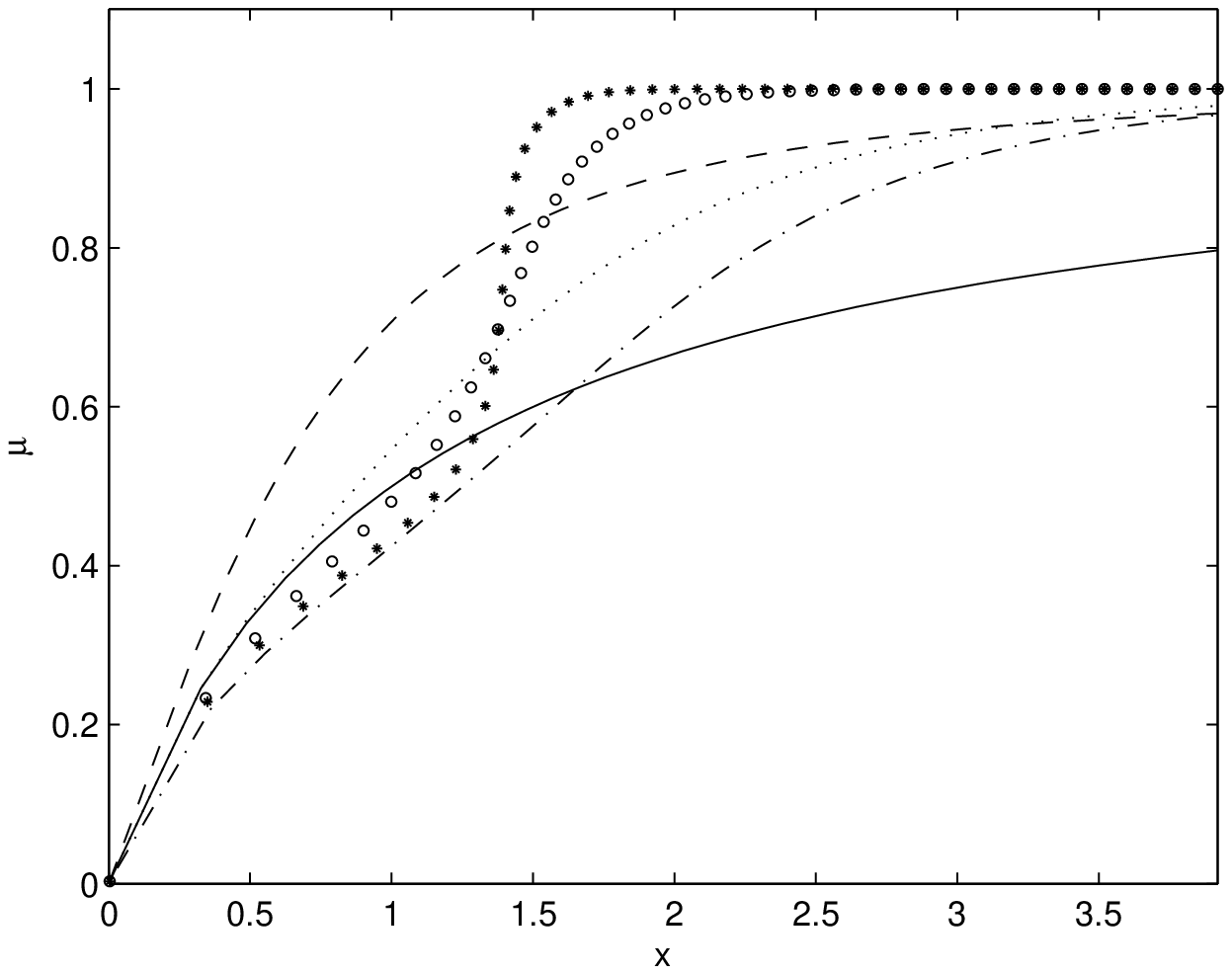} &
\includegraphics[width=0.5\columnwidth]{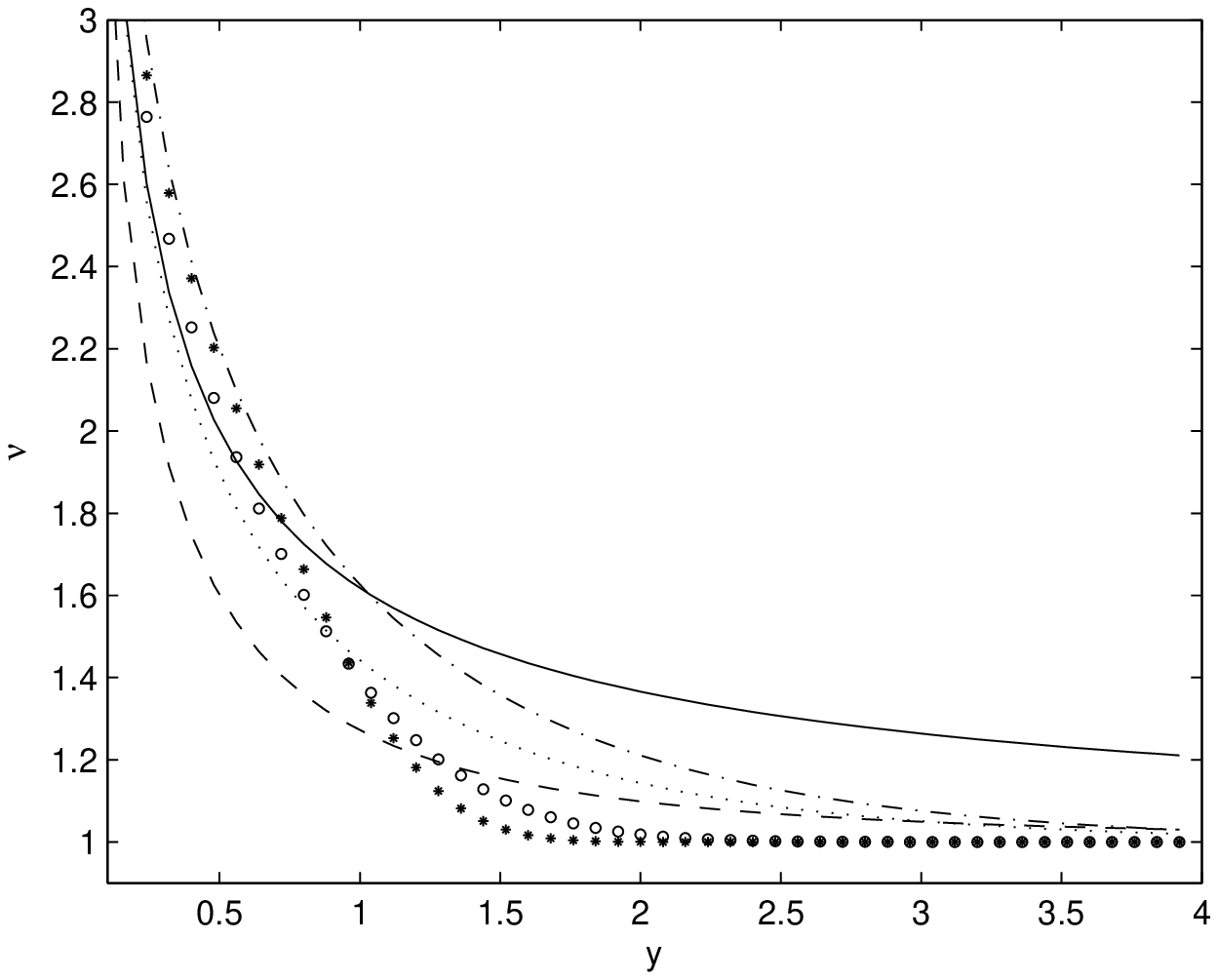}\\
\includegraphics[width=0.5\columnwidth]{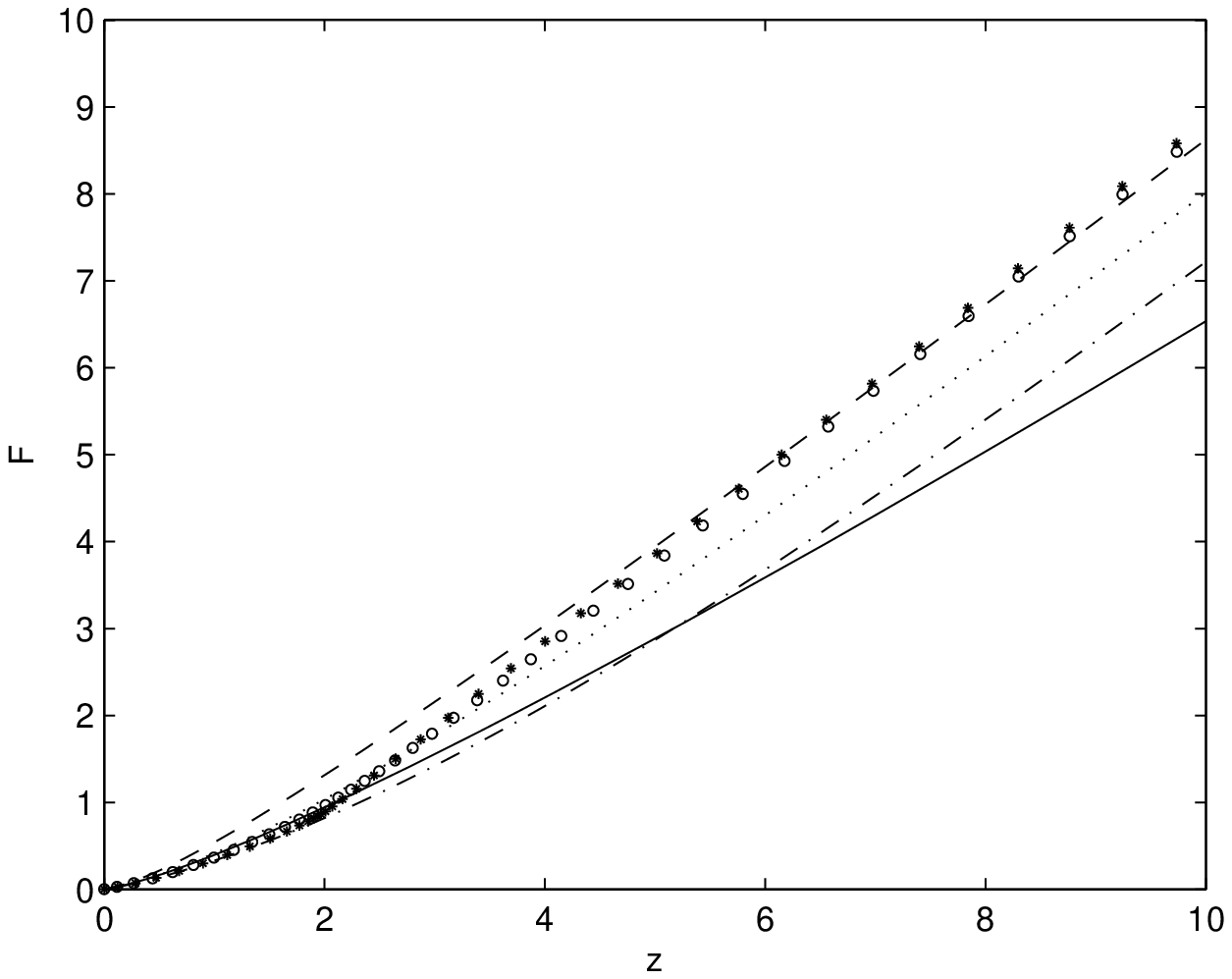}\\
\end{tabular}
\caption{The form of $\mu(x)$, $\nu(y)$, and the Lagrangian function
$F(z)$  for $\mu_1$ (solid), $\mu_2$ (dashed), $\tilde\mu_{0.5}$
(dotted) $\tilde\mu_1$ (dash-dot), $\bar\mu_{2}$ (circles) and
$\bar\mu_3$ (stars).} \label{fig1}
\end{figure}

We show in Fig. \ref{fig1} the forms of $\mu, ~\nu$, and the
Lagrangian function $F(z)$ for two choices of the parameter in
each family. We see that for ${\tilde\mu}_\a$ and ${\bar\mu}_\a$
the approach to 1 can be as gradual as ${\mu}_1$ in the MOND
regime, but then with a rather more rapid transition when
$a/\az>1$.

We can classify the interpolating functions according to several
attributes: how fast $\mu(x)$ approaches 1 for very high values of
$x$ (this is important for calculating the MOND correction in highly
Newtonian systems, such as the solar system, and is not crucial
here); how far $\mu(x)$ stays near 1 as $x$ approaches $x=1$ from
above; how fast $\mu(x)$ departs from the initial $\mu\sim x$ at low
values of $x$. All of the attributes affect the appearance of the
ring at $r_t$, and within each of the one-parameter families it is
that one parameter that controls all aspects. To be able to explore
the different attributes separately we consider these three
families.

\begin{figure}
\begin{tabular}{rl}
\tabularnewline
\includegraphics[width=0.5\columnwidth]{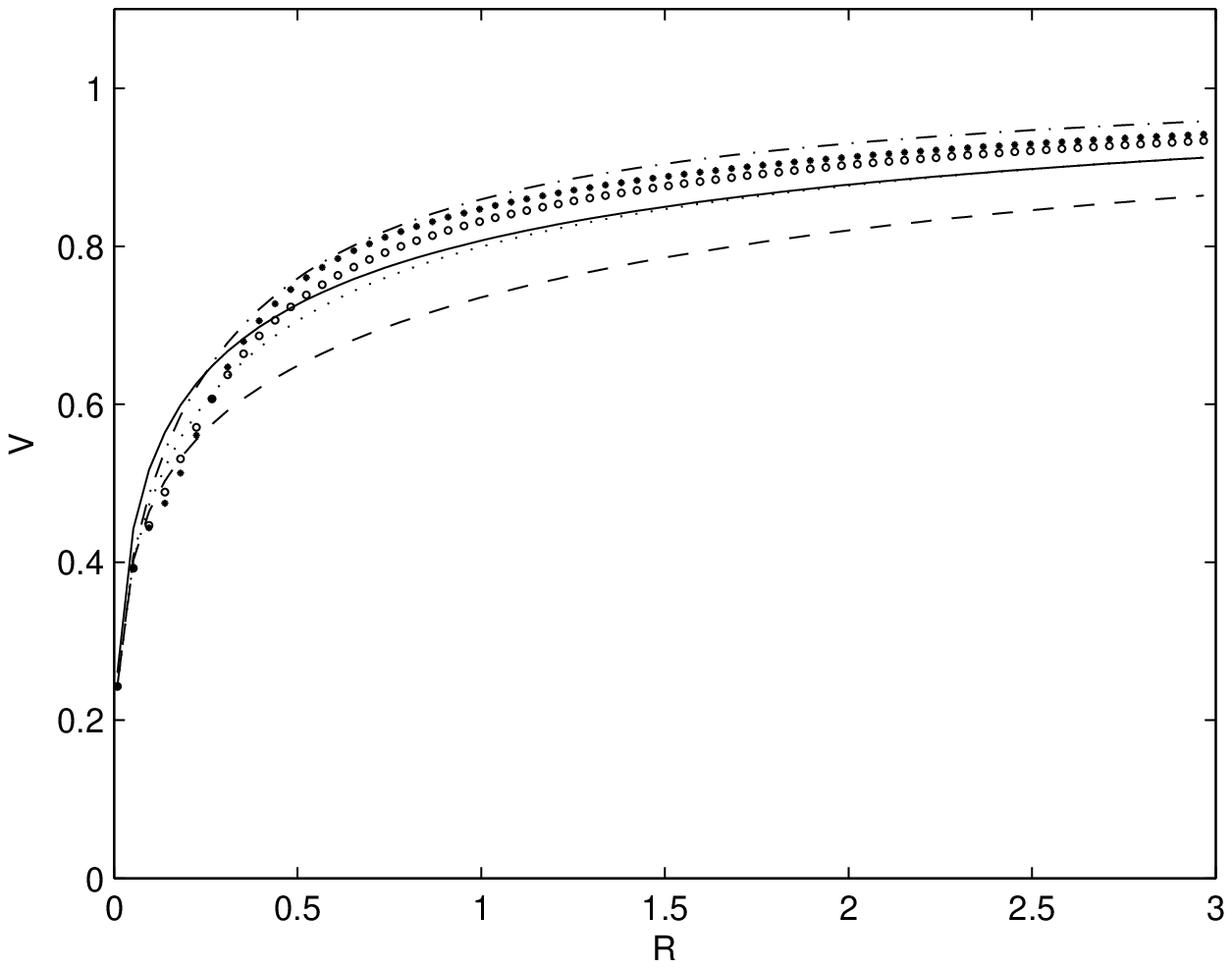}  &
\includegraphics[width=0.5\columnwidth]{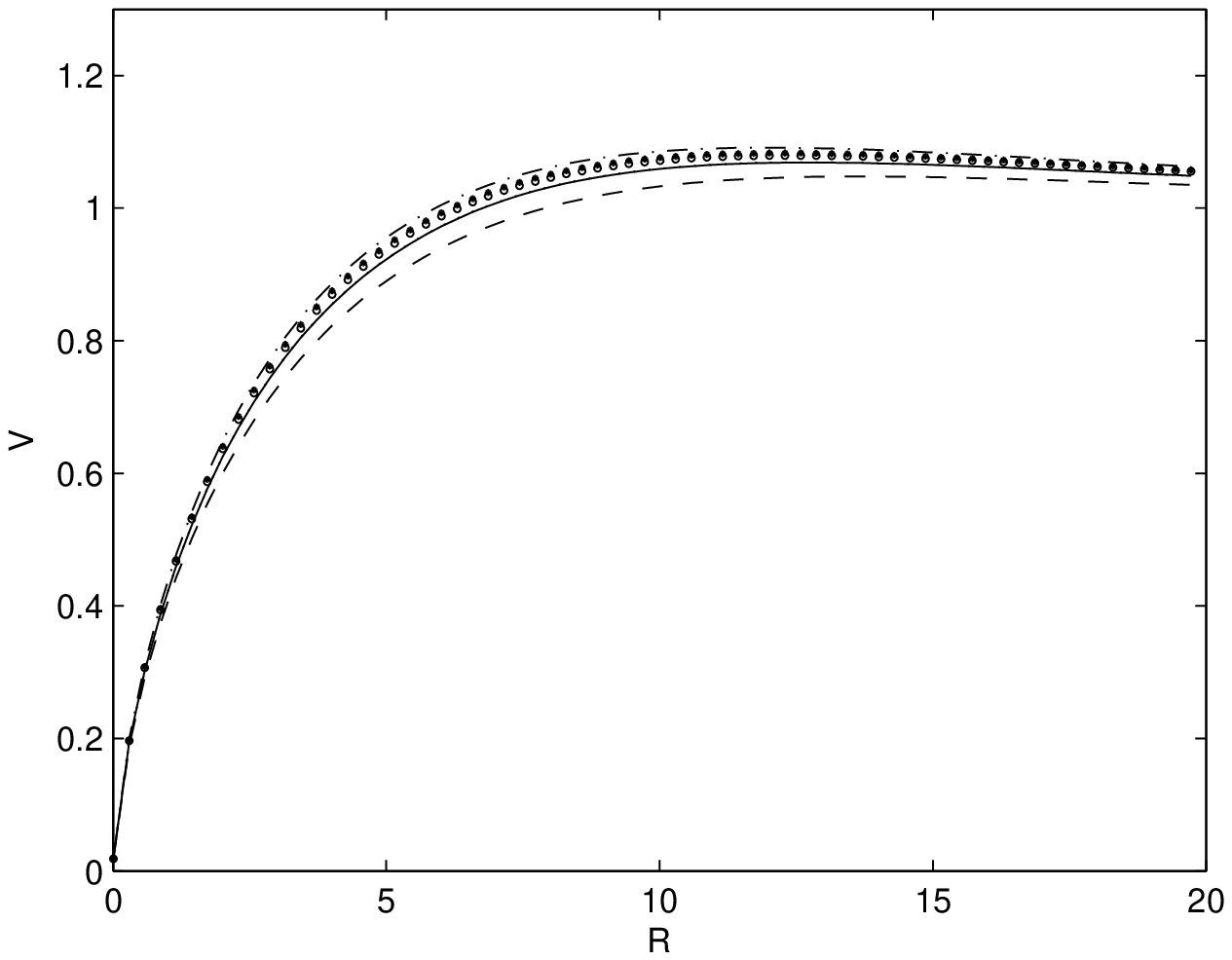}\\
\includegraphics[width=0.5\columnwidth]{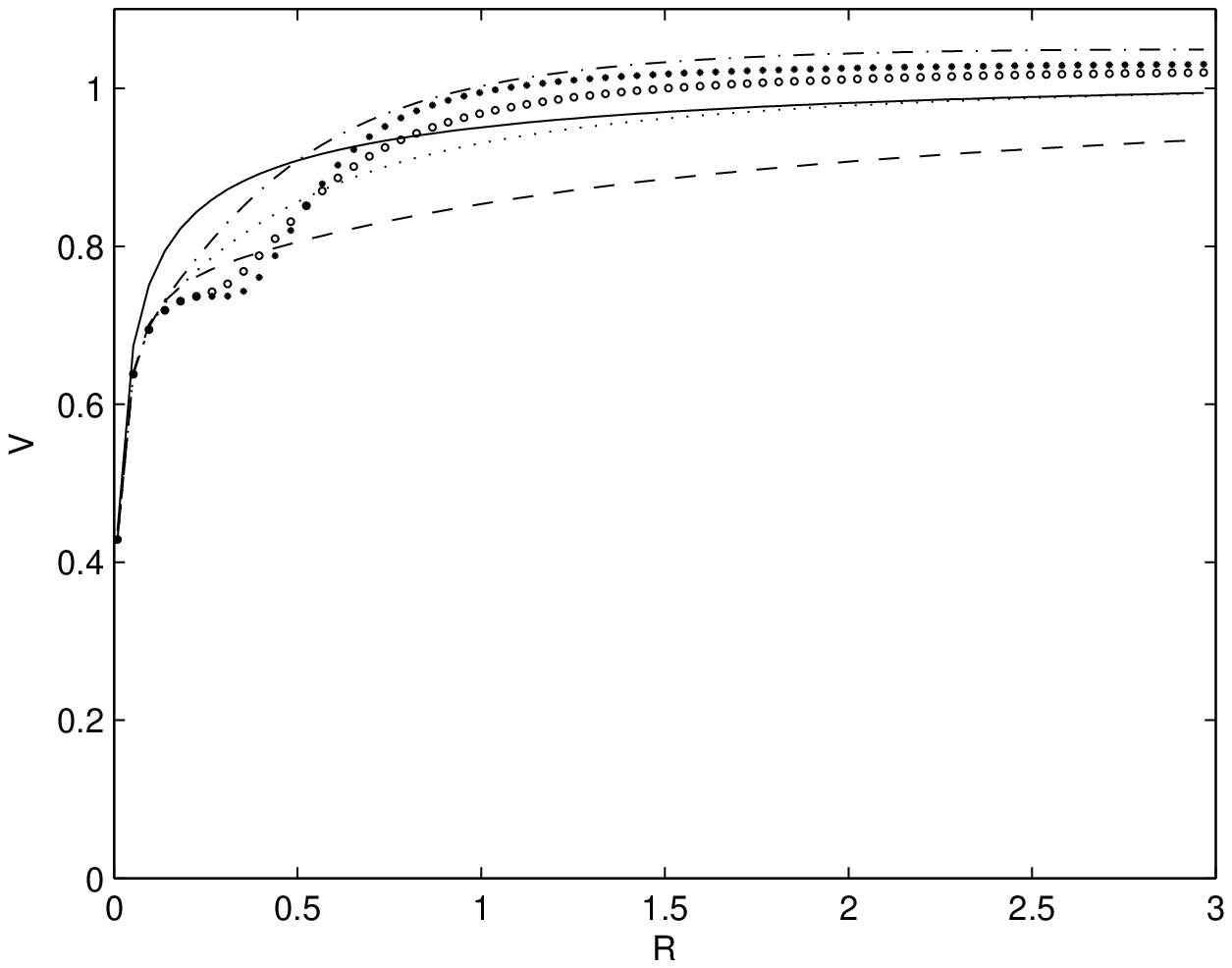}  &
\includegraphics[width=0.5\columnwidth]{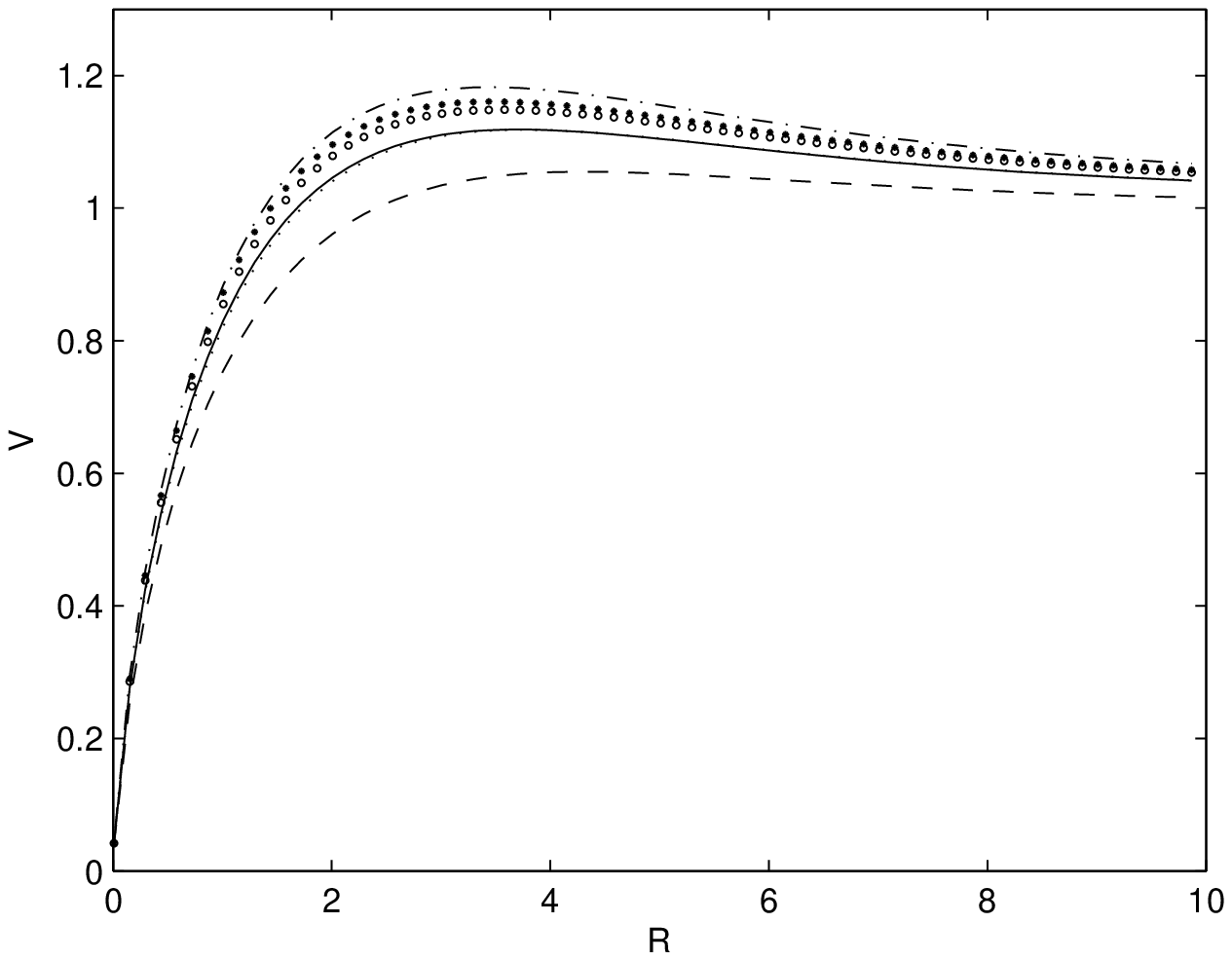}\\
\includegraphics[width=0.5\columnwidth]{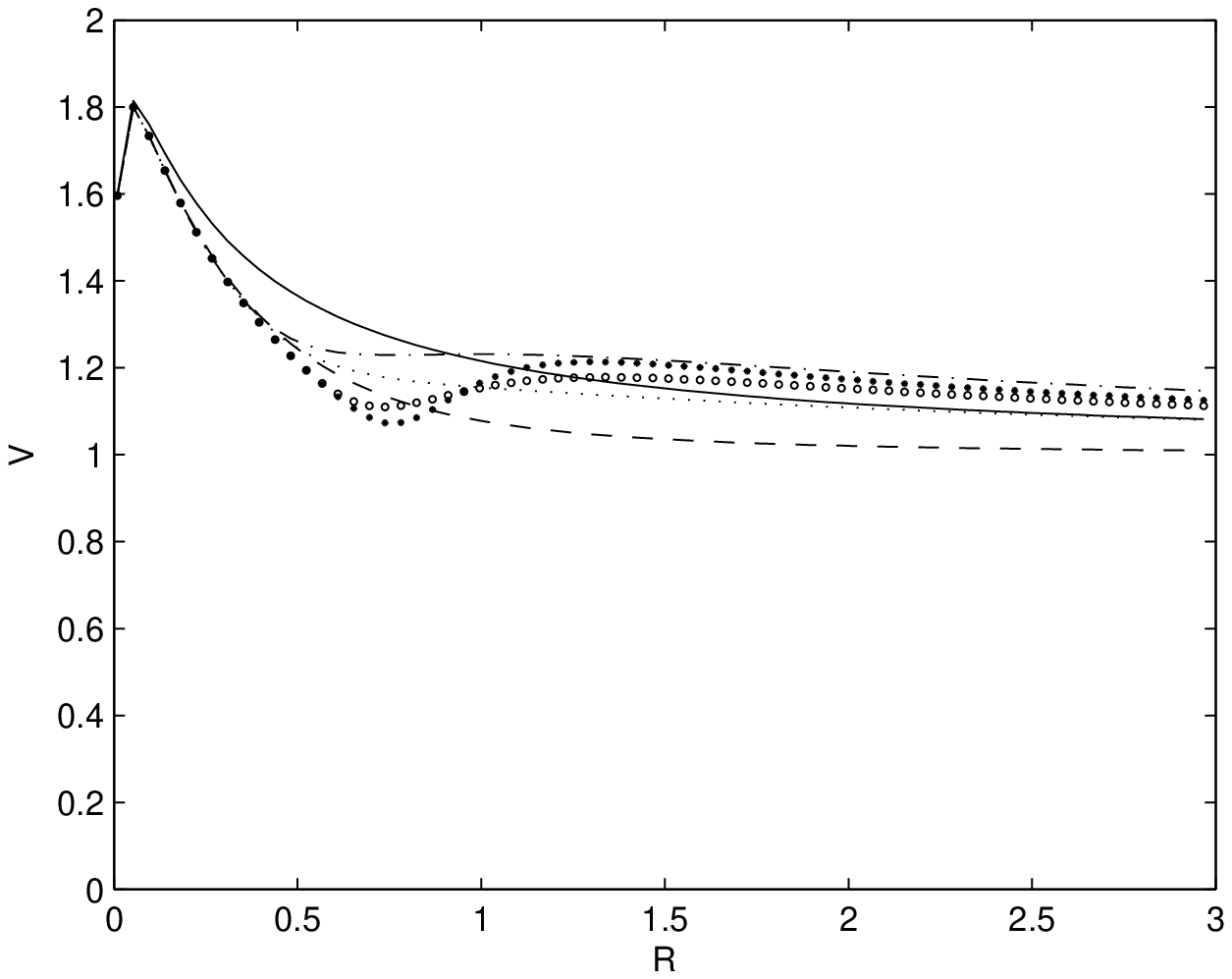}  &
\includegraphics[width=0.5\columnwidth]{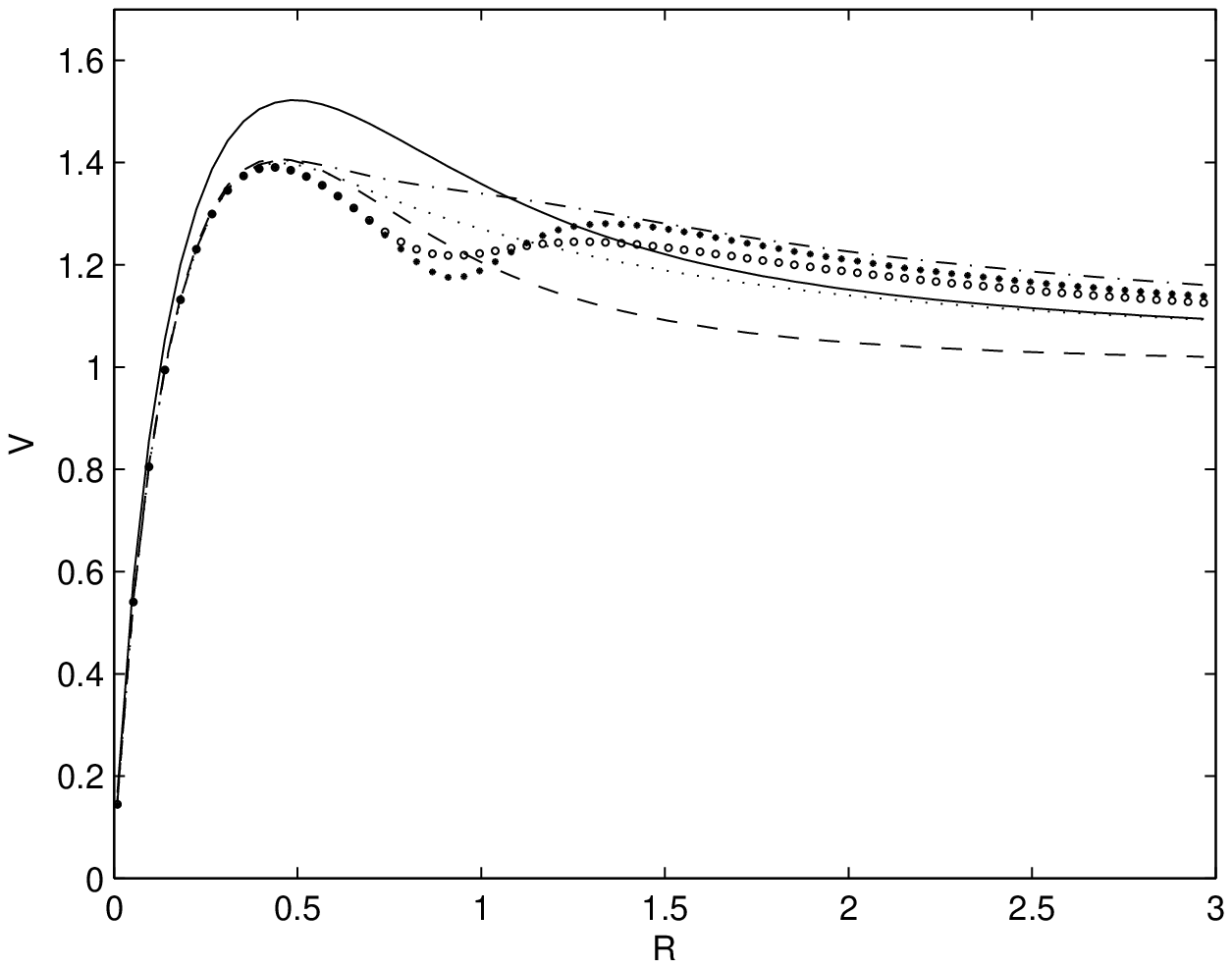}\\
\end{tabular}
\caption{MOND rotation curves for de Vaucouleurs spheres (left) and
exponential discs (right). The former (from top to bottom) for
effective radii 2, 1, 1/6  (all in units of the transition radius);
the latter (from top to bottom) for scale lengths 3, 1, 1/5 . The
different curves are for different interpolating functions: $\mu_1$
(solid), $\mu_2$ (dashed), $\tilde\mu_{0.5}$ (dotted) $\tilde\mu_1$
(dash-dot), $\bar\mu_{2}$ (circles) and $\bar\mu_3$ (stars).}
 \label{fig2}
\end{figure}

To demonstrate the ability of rotation curve analysis to
discriminate between the interpolating functions, we show in Fig.
\ref{fig2} the rotation curves for a pure exponential disc and pure
de Vaucouleurs sphere of different scale lengths and for different
choices of the interpolating function. We see that with the more
extreme forms in our sample one even expects to see an orphan
feature directly in the rotation curve of galaxies with high enough
surface densities, or mean acceleration. (An orphan feature is one
that is not directly dictated by the underlying source distribution
and hence does not appear on the Newtonian rotation curve.) A more
systematic study of this point is required to insure that we adopt a
form of $\mu$ that is compatible with rotation curve analysis (and
see section 5 for a beginning of such analysis). We reemphasize,
however, that the forms of $\mu$ we use here are not suggested as
real candidates. We chose them essentially to span some range of
behaviors so that we can demonstrate the dependence of the ring
phenomenon on $\mu$, although some of the new forms of $\mu$ do seem
to do better for high acceleration galaxies (see section 5) as well
as being more favorable for the visibility of the "ring".

\subsection{Structure along the line of sight}

The thin-lens approximation, so useful in conventional dynamics,
does not apply in MOND : Because of the inherent nonlinearity of
MOND, even in the nonrelativistic limit, two (``baryonic'') mass
distributions having the same projected surface density
distributions for a given line of sight, do not, in general, have
the same surface densities of the corresponding PDM (Mortlock and
Turner 2001, Milgrom 2002). For example, as explained in Milgrom
(2001) $N$ equal masses well separated along the line of sight
produce, in the deep MOND regime,  a deflection angle, for a given
impact parameter, that is $\sqrt{N}$ times larger than when the $N$
masses are melded in one. Such effects will be clearly shown by our
numerical results below.

It should be emphasized that for gravitational lensing in a theory
such as TeVeS, the relation between the deflection of photons and
the total weak field force is the same as that in General
Relativity.  This means that the surface density derived for the PDM
(using, for example, the lens convergence estimated from the
observed shear) is identical to that obtained by integrating the
density of the PDM along the line of sight.

The discussion above eq.(\ref{xxi}) tells us that for two mass
distributions of the same total (true) mass we have
 \beq \int_0^R[\S_1(R')-\S_2(R')]R'^2~dR'\rar 0, \label{xxiii}\eeq
 in the limit $R\rar\infty$.

\section{Examples}

\subsection{Spherical lenses}
Here we discuss the simple case of a spherical lens represented by a
point mass; so that our results can be applied directly everywhere
outside any spherical mass. In this case, since $g_N=r^{-2}$, we
have
 \beq 4\pi\rho_p=r^{-2}{d(r^2g)\over dr}=r^{-2}{d[\nu(r^{-2})]\over
 dr}=-2\nu'(r^{-2})r^{-5}. \label{x}\eeq

\begin{figure}
\begin{tabular}{rl}
\tabularnewline
\includegraphics[width=0.5\columnwidth]{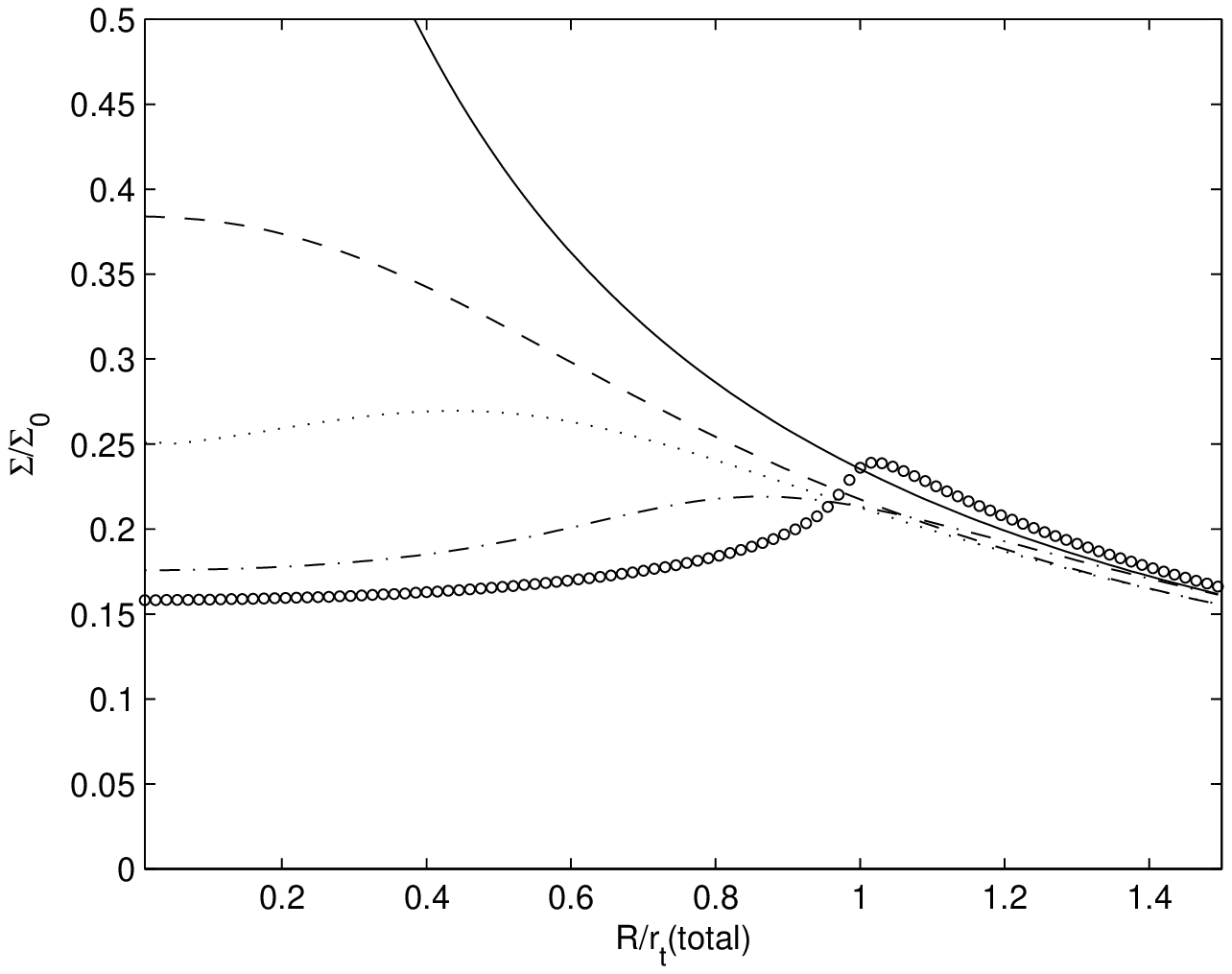} &
\includegraphics[width=0.5\columnwidth]{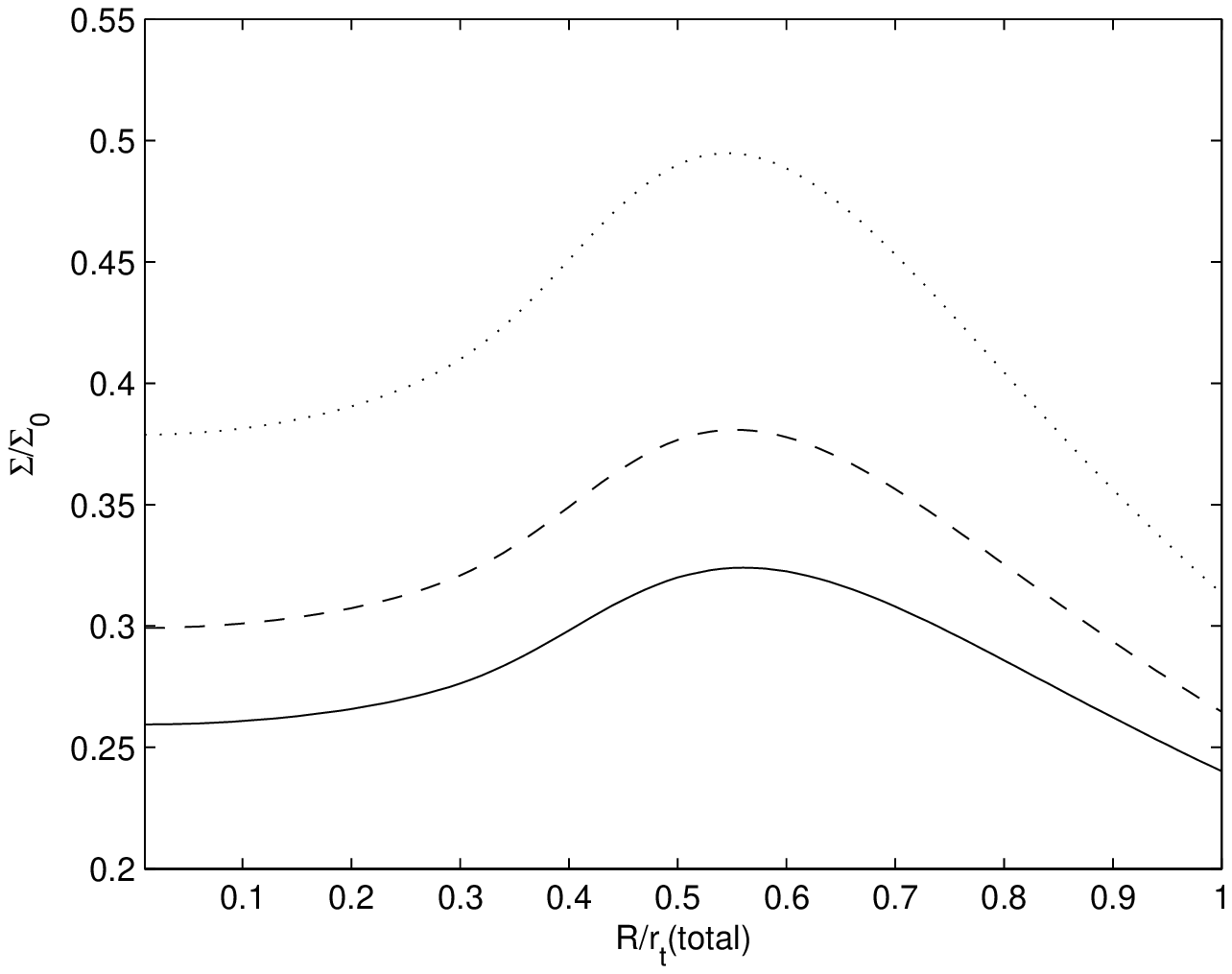}\\
\includegraphics[width=0.5\columnwidth]{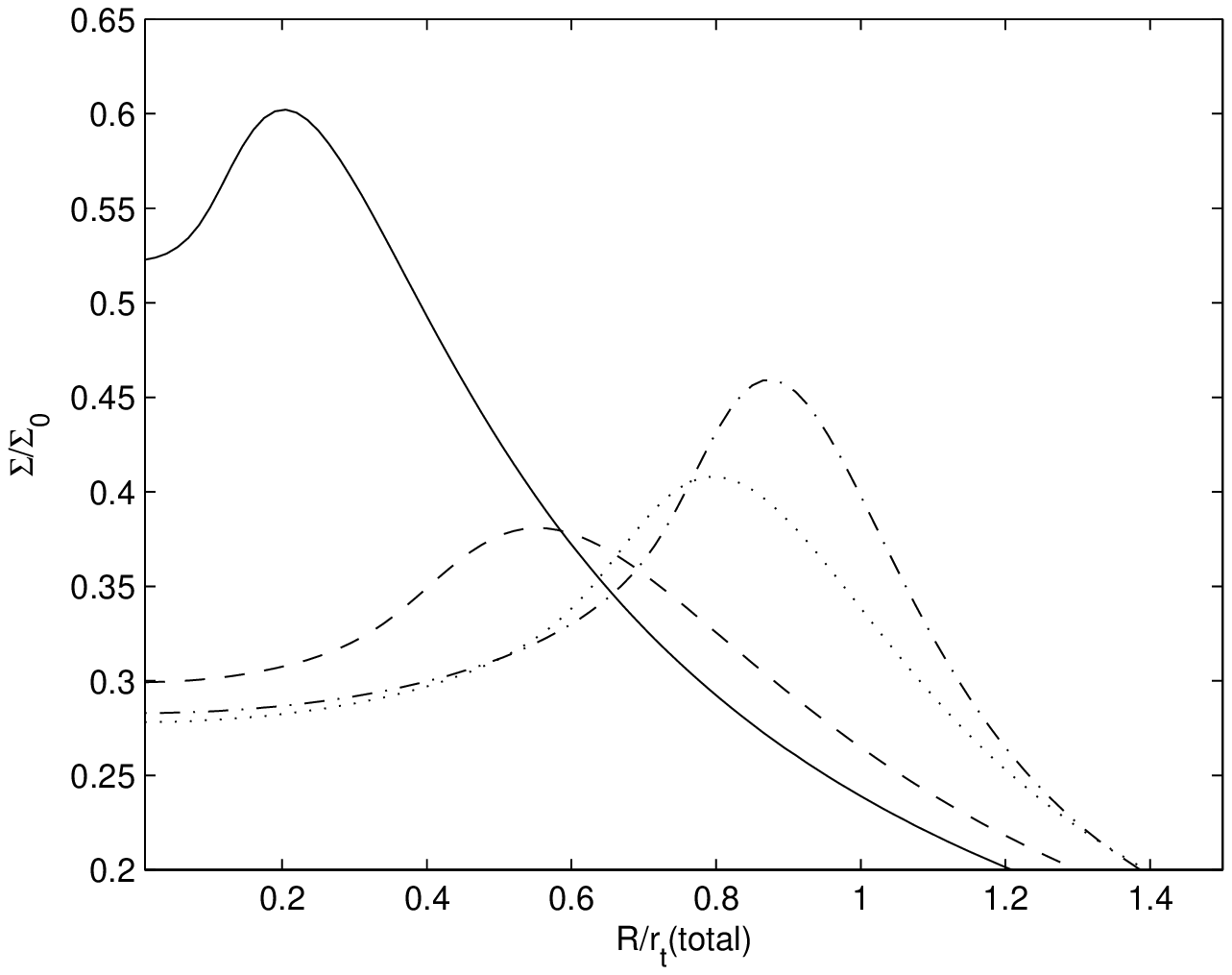}\\
\end{tabular}
\caption{The MOND predicted PDM surface density, in units of
$\Sz$, for a point mass with different forms of the MOND function.
Top left: $\mu_{\a}$ with $\a=1,~3/2,~2,~4,~50$ (the higher $\a$
the lower $\Sigma $ at low radii). Right: $\tilde\mu_{\a}$ with
$\a=0.25,~0.5,~1$ (the higher $\a$ the higher $\Sigma$). Bottom
left: $\bar\mu_{\a}$ with $\a=0.5$ (solid), $\a=1$ (dashed),
$\a=2$ (dotted), $\a=3$ (dot-dash).} \label{fig3}
\end{figure}

We show in Fig. \ref{fig3} the surface density of the PDM for a
point mass with different choices of $\mu$. We see that the
occurrence of a maximum leading to the appearance of a ring is
rather generic. Of all the choices of $\mu$ shown in the figure  a
maximum at a finite radius does not appear only for $\mu_1$ and
$\mu_{3/2}$. This is easy to understand: For this form of $\mu$ we
have in the high acceleration limit $g\approx
g_N+\a^{-1}g_N^{(1-\a)}$. Since $g_N\propto r^{-2}$, we have
$\rho_p\propto r^{2\a-3}$. So for $\a\le 3/2$ there is no ``hole''
in the PDM density, and a maximum can appear only for $\a>3/2$. We
also see that the radius (in units of $r_t$), the height, and the
contrast of the maximum depend rather sensitively on the properties
of the interpolating function. The location of the peak is
controlled mainly by the approximate $x$ value at which $\mu(x)$
approaches 1 (in the vicinity of $x=1$, not in the far asymptotic
regime). So, for example, the limiting case $\mu_{\infty}$ produces
a maximum at $r=1$, which is the largest for all the functions we
study here. For the other forms of $\mu$ we use it goes near a value
of 1 for $x>1$; accordingly, the ring occurs at $r<r_t$. The height
of the peak is sensitive to how low $\mu(x)$ stays for values of
$x<1$. With this guidelines one can construct $\mu$ forms to
engineer the ring properties. To recapitulate, the formation of the
apparent ring is generic if the mass is centrally condensed with
respect to $r_t$. It is not produced only for some special forms of
$\mu$ such as ${\mu}_\a$ with $\a \le 3/2$.

\subsection{Aligned dumbbell lenses}

\begin{figure}
\begin{tabular}{rl}
\tabularnewline
\includegraphics[width=0.5\columnwidth]{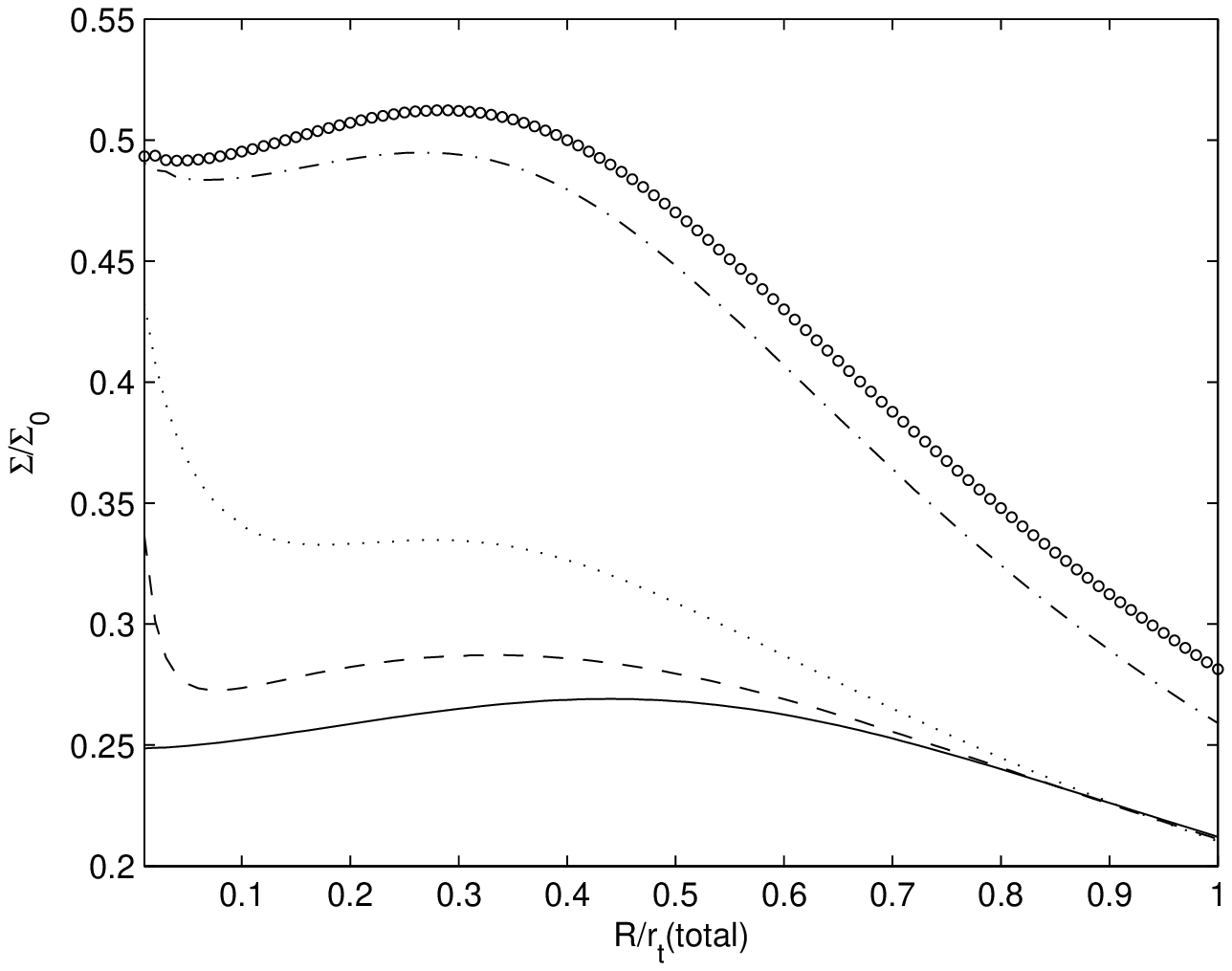} &
\includegraphics[width=0.5\columnwidth]{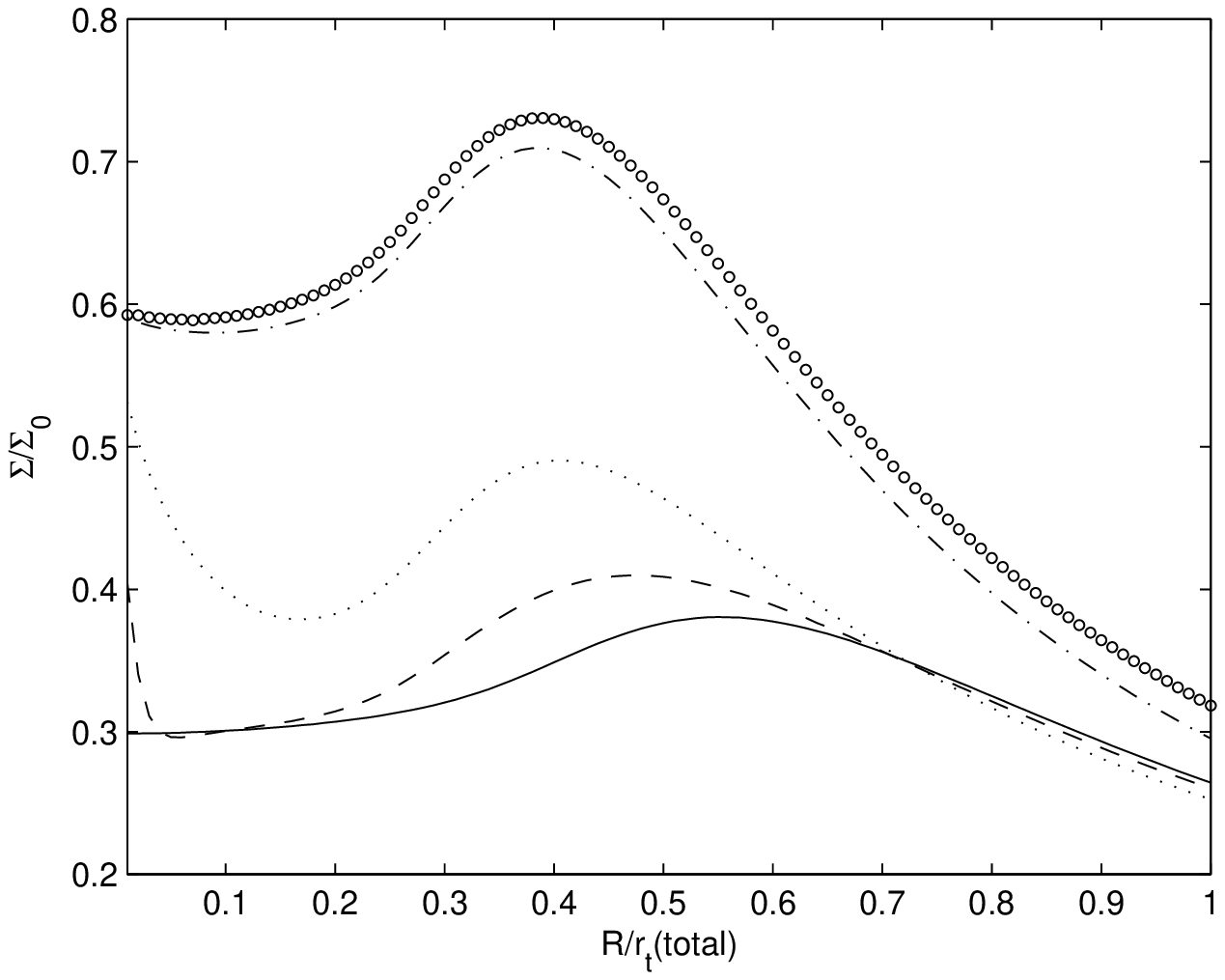}\\
\includegraphics[width=0.5\columnwidth]{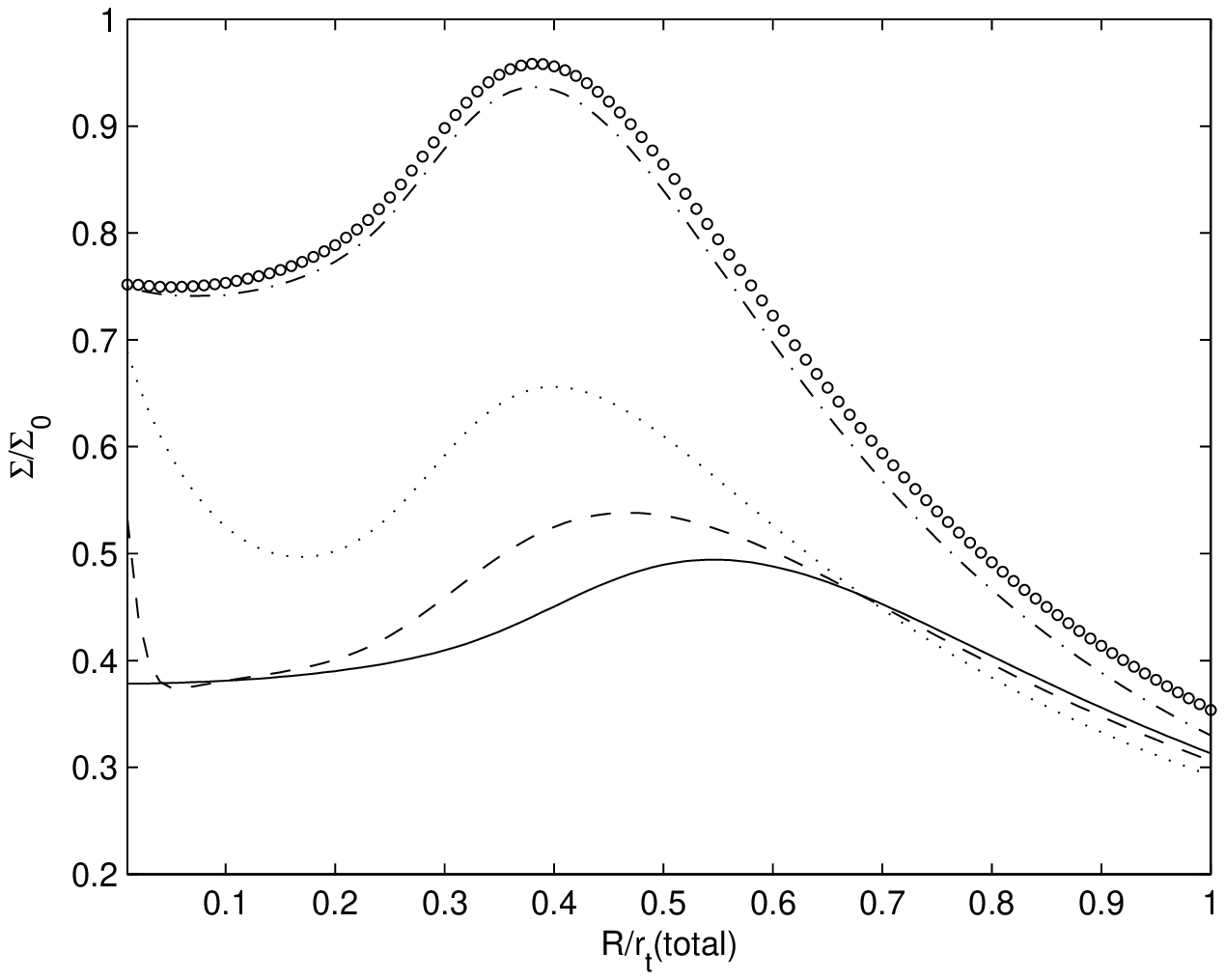} &
\includegraphics[width=0.5\columnwidth]{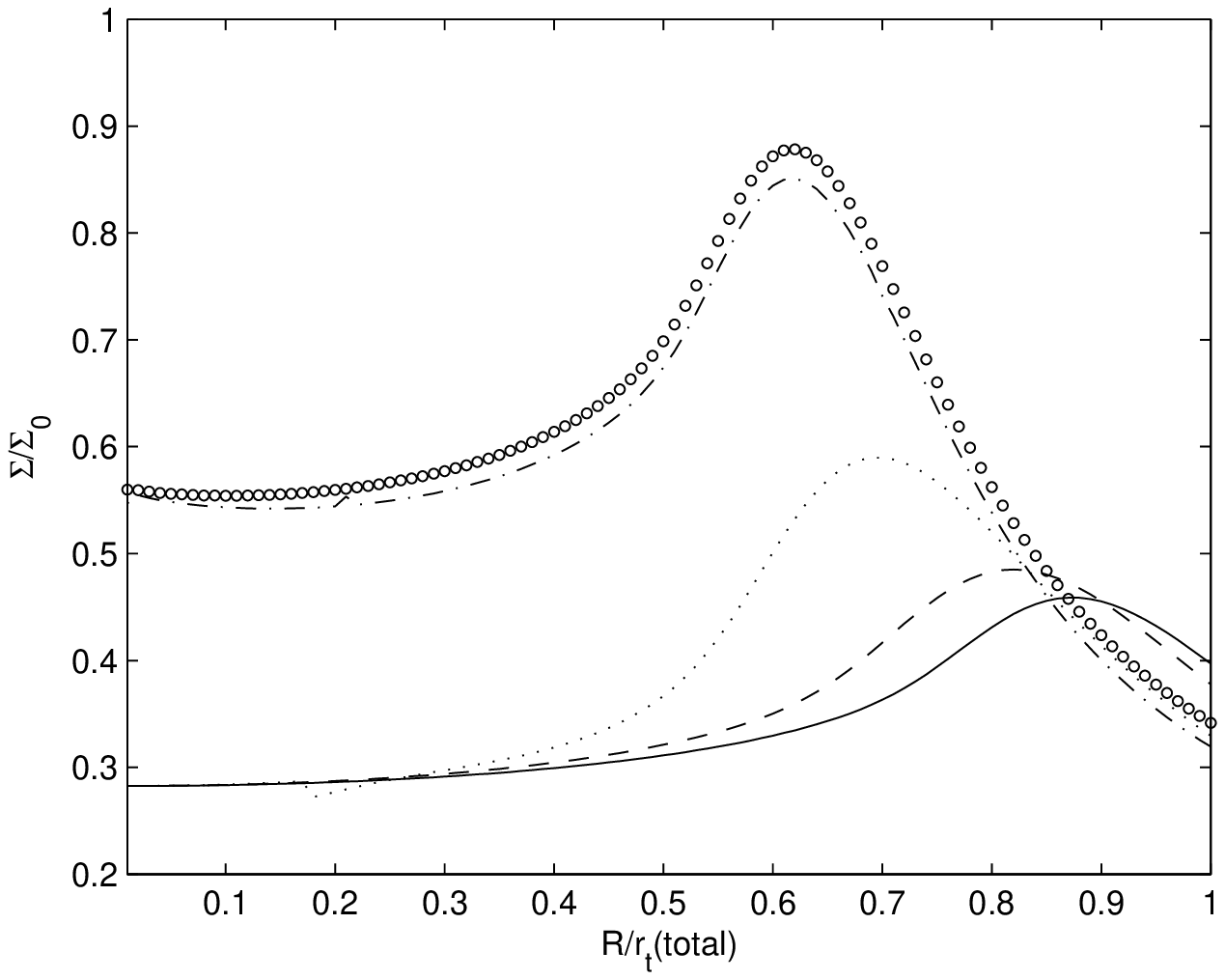}\\

\end{tabular}
\caption{The MOND predicted PDM surface density, in units of
$\Sz$, for two equal point masses along the line of sight
separated by 0 (solid), 0.5 (dashed), 1 (dotted), 4
(dashed-dotted), 20 (circles) transition radii for the total mass.
For $\mu_2$ (upper left), $\tilde\mu_{0.5}$ (upper right),
$\tilde\mu_{1}$ (lower left), and $\bar\mu_3$ (lower right).}
\label{fig4}
\end{figure}

Here, for simplicity, we still use the algebraic relation,
eqs.(\ref{ii})(\ref{iia}), even though it is not exact anymore. We
show in  Fig. \ref{fig4} the predicted surface density of the PDM
for two equal point masses aligned with the line of sight for
different separations, and different choices of the interpolating
function. Figure \ref{fig4} [together with Fig.(\ref{fig5})]
demonstrates clearly that the thin-lens approximation is not valid
in MOND as all the mass models it describes have the same projected
(baryonic) mass. They all give the same result at radii much larger
than the separation (where they all act like a point mass); but at
smaller radii elongation of the mass along the line of sight
enhances the lensing signal (see Milgrom 2002). When the separation
is much smaller than 1 (the individual transition radius) the system
acts like one point mass of value 2, whose transition radius is thus
$2^{1/2}$; however, as regards the normalization, the resulting
surface density of the PDM is the same as that of a single mass (as
this does not depend on the mass). In the other extreme, when the
separation is much larger than 1, the PDM surface density is simply
double that of a single mass in magnitude, and it has the same
radial distribution. This is also in line with relation
(\ref{xxiii}).

For brevity's sake we show only results for equal masses. Clearly,
with unequal masses a more complex profile results, with possibly
two maxima.

Note that for some separations and some forms of $\mu$, the PDM
surface density profile shows an upturn towards small radii. This is
another orphan feature of MOND connected with the vicinity of the
point where $g=0$.

\subsection{A thin mass rod along the line of sight}

 Consider a mass $M$ of constant density, with
diameter $D$ and length $L$ aligned with the line of sight. Assume
that $D/2\ll r_t= (MG/\az)^{1/2}$ (we shall take $D=0$). For there
to be an enhanced effect over a point mass we have to have $L\gg
r_t$, then the mass will act roughly like
$N=L/r_t=(MG/L^2\az)^{-1/2}$ separate masses each of value $m\equiv
M/N$. The effect of increasing $N$ is thus to bring closer the
radius of the maximum and at the same time raise its height, also in
line with eq.(\ref{xxiii}).

Again, taking as an example the $\mu_{\a}$ family, in the high
acceleration limit we have $g\approx g_N+\a^{-1}g_N^{(1-\a)}$. For
small $N$ we are basically still in the point mass limit. But for
large N, when we are near the mass, the field is approximately that
of an infinite wire; so $g_N\propto r^{-1}$. We then have
$\rho_p\propto r^{\a-2}$; so, now the limiting value of $\a$ (for
getting a "hole") for large $N$ is $\a=2$.

\begin{figure}
\begin{tabular}{rl}
\tabularnewline
\includegraphics[width=0.5\columnwidth]{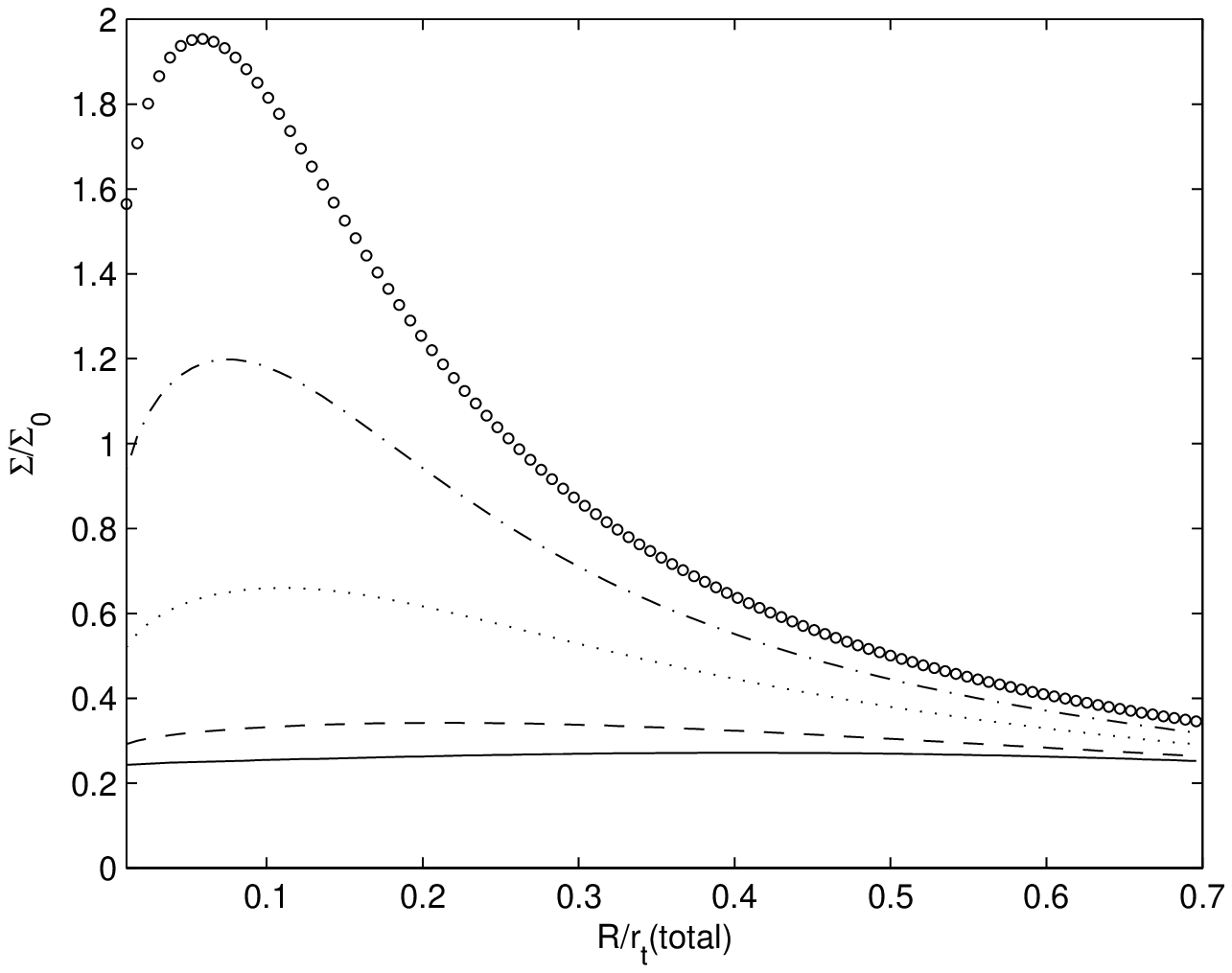}  &
\includegraphics[width=0.5\columnwidth]{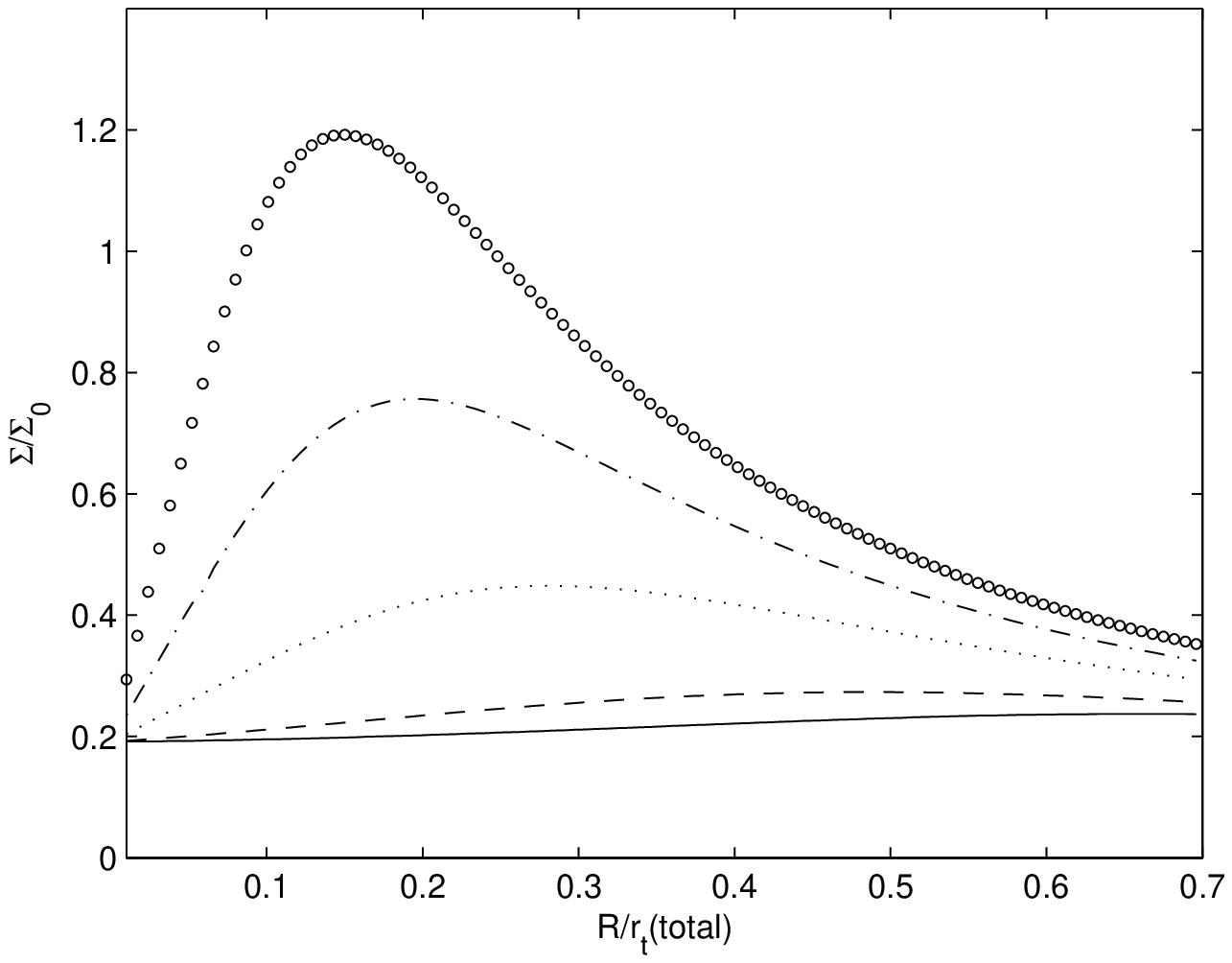}\\
\includegraphics[width=0.5\columnwidth]{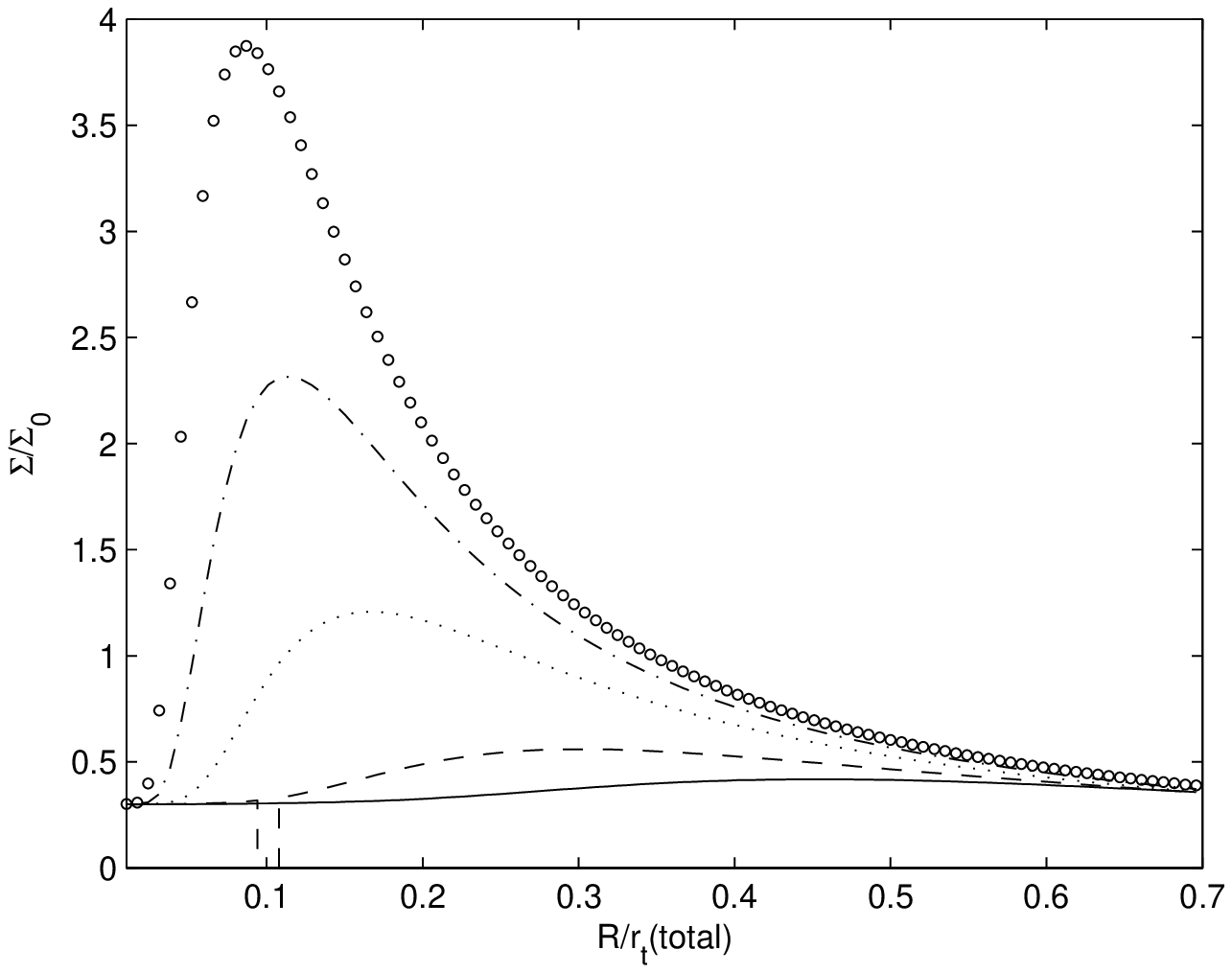}  &
\includegraphics[width=0.5\columnwidth]{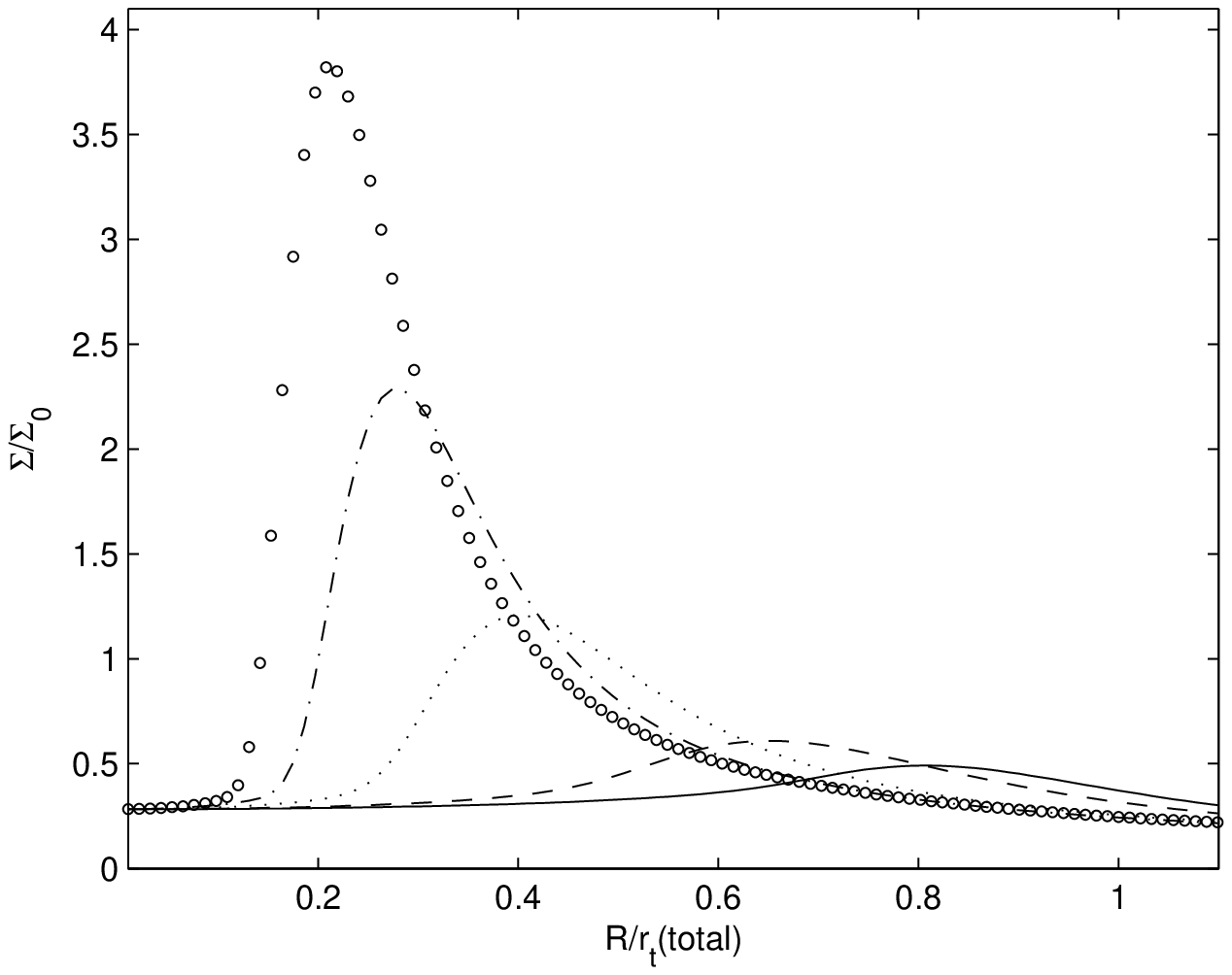}\\
\end{tabular}
\caption{The MOND predicted PDM surface density, in units of
$\Sz$, for a line mass of different lengths $N=1,~2,~4,~6,~8$ (in
units of the transition radius; higher peak $\Sigma$ for higher
$N$). Different interpolating functions are used: $\mu_{2.2}$ (top
left), $\mu_3$ (top right), $\tilde\mu_{0.5}$ (bottom left), and
$\bar\mu_{3}$ (bottom right). } \label{fig5}
\end{figure}

We show in Fig. \ref{fig5} the projected phantom surface density
for different values of $N$ and for various interpolating
functions.

\section{Possible relevance to Cl 0024+17}

 Jee et al. (2007) offer an explanation of the apparent ring in Cl
0024+17 in terms of an actual ring of dark matter, in 3-D, produced
by a past collision of two clusters occurring along the line of
sight. The two clusters are now separated from each other. Jee et
al. adduce cogent evidence to show that the cluster is indeed
composed of (at least) two separate structures along the line of
sight (bi-modality of the galaxy velocity distribution, and a
factor-of-2 discrepancy between the masses deduced from X-ray
analysis and from weak lensing). Famaey et al. (2007a) give a
similar explanation in the framework of MOND building on the fact
that MOND too requires DM in the cores of clusters. The past
collision is a necessary ingredient in these explanations.  Our
mechanism depends only on the present mass distribution and cares
little as to whether or not there occurred a collision in the
system. (Although in some unknown way the collision could have
created a special mass distribution that is expedient to a ring
formation.)

 As stated above, galaxy clusters are required to contain
large quantities of as yet undetected matter, even in the context of
MOND. Direct observations thus do not tell us how much conventional
(MOND source) matter there is in the cluster and how it is
distributed. In addition, as we explained above, the actual
distribution of that mass in 3-D is very crucial, and this can
certainly not be deduced from the observations. For these reasons,
and because we do not know the exact form of the interpolating
function, we are not in the position to construct a specific model
for the ring in Cl 0024+17 (Jee et al. also do not have a specific
model only a plausibility scenario, as the required details of the
underlying collision are not known). Instead, we want to see only if
the observed parameters of the ring can naturally result from the
mechanism that we discuss here.

The surface density distribution deduced by Jee et al. (2007) is
unusual for single clusters: It is characterized by a very
concentrated component followed by an almost constant surface
density out to the last measured point; the ring appears on the
background of this plateau. As stated above it is the fact that
there is a central concentrated component that is conducive to ring
appearance.

There are four observables that we want to reproduce in rough terms:
the radius at the ring maximum, the mass within that radius, the
surface density at the location of the maximum, and the contrast of
the ring. Jee et al. give the radius of the ring (presumably that of
the maximum) as $400\kpc$ (75"). According to our previous results
this should be some fraction of the transition radius. So $r_t$
would be typically between $500 \kpc$ (if $r/r_t=0.8$) and
$1600\kpc$ if ($r/r_t=.25$). We have

\beq r_t=833(M/5\times
10^{14}\msun)^{1/2}(\az/10^{-8}\cmst)^{-1/2}\kpc, \label{xx}\eeq so
the required (projected) mass inside the ring would be somewhere
between $2\times 10^{14}\msun$ and $20\times 10^{14}\msun$. This are
only rough estimates but they do define the right ballpark, as the
projected dynamical mass within the radius of the maximum, according
to Fig. 12 of Jee et al.,  is $\sim 8\times 10^{14}\msun$ and that
within the observed minimum (at 50") is $\sim 4\times 10^{14}\msun$.

 The critical surface density with respect to
which Jee et al. plot their deduced surface densities: $\Sigma_c =
c^2(4GD)^{-1} = 0.35\gcmt(D/1 \gpc)^{-1}$, where $D =
D_dD_{ds}/D_s$; the $D$s are angular diameter distances to the lens
(d), the source (s), and between the two (ds). Jee et al. use for
the fiducial red-shift of the source  $z=3$. We take $D_{ds}\sim
D_s$, and deduce $D_d$ by noting that they quote the radius of the
ring as
 $400\kpc=75"$. This gives $D\approx
1.1 \gpc$. So $\Sigma_c \sim 0.32\gcmt$. Since $\Sz\sim 0.15\gcmt$
(for $\az=10^{-8}\cmst$), $\Sigma_c\approx 2\Sz$. Jee et al. find at
the location of the ring  $\Sigma\sim 0.7\Sigma_c\sim 0.22 \gcmt\sim
1.5\Sz$.  From Fig. 10 of Jee et al. we read for the ring contrast a
difference between minimum and maximum of $\sim 0.06\Sz$. This is
rather easy to achieve.

\begin{figure}
\begin{tabular}{rl}
\tabularnewline
\includegraphics[width=0.5\columnwidth]{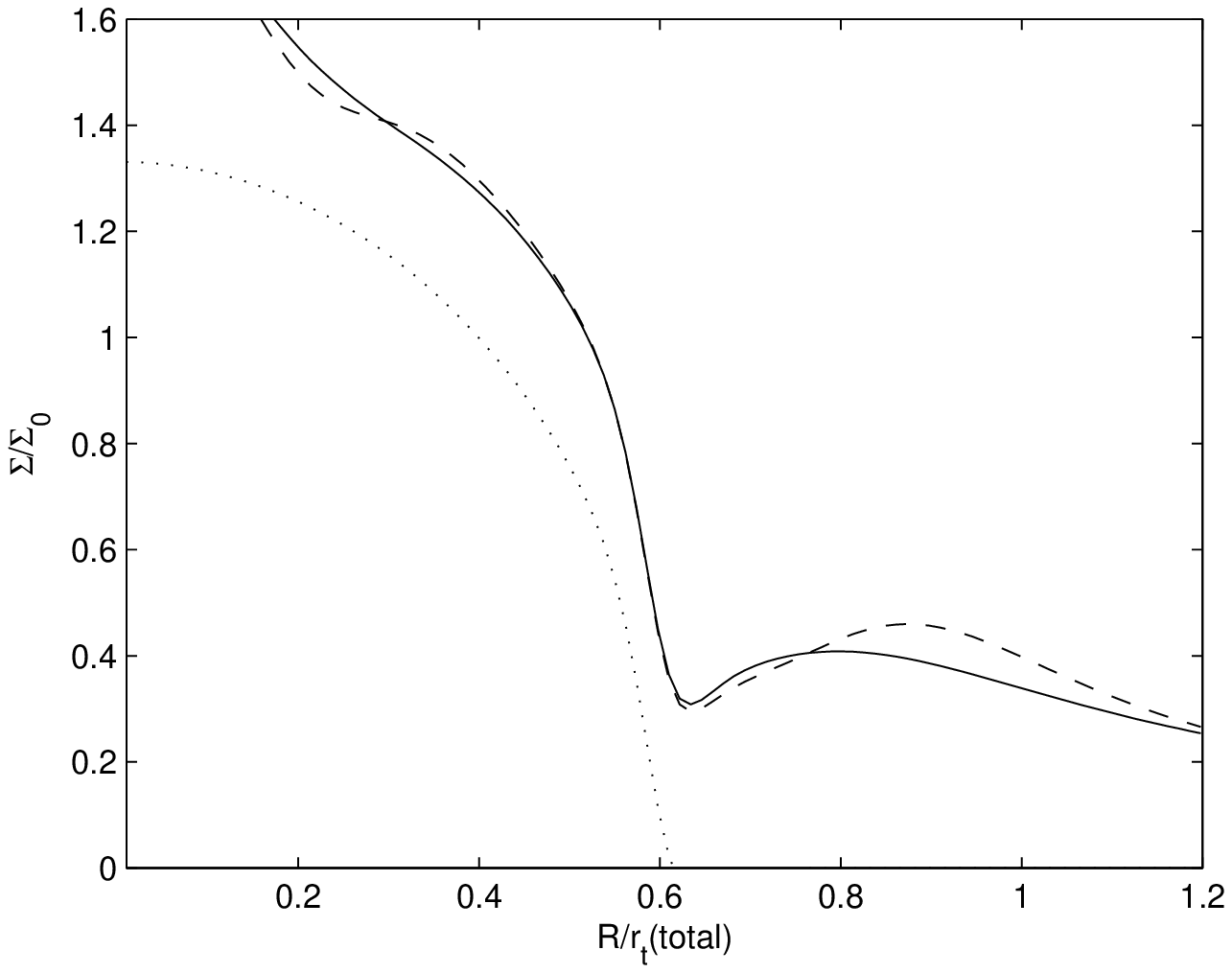} &
\includegraphics[width=0.5\columnwidth]{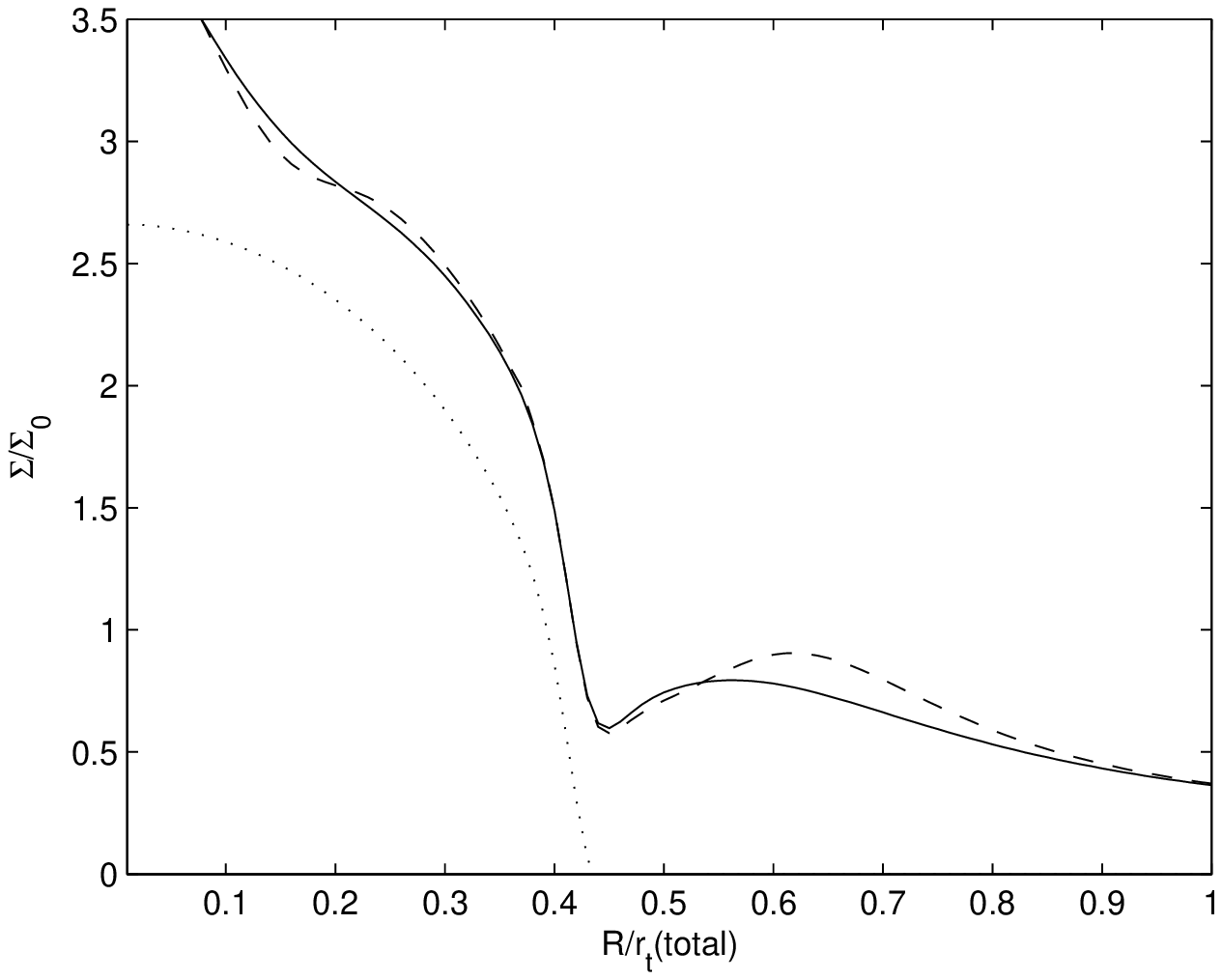}\\
\includegraphics[width=0.5\columnwidth]{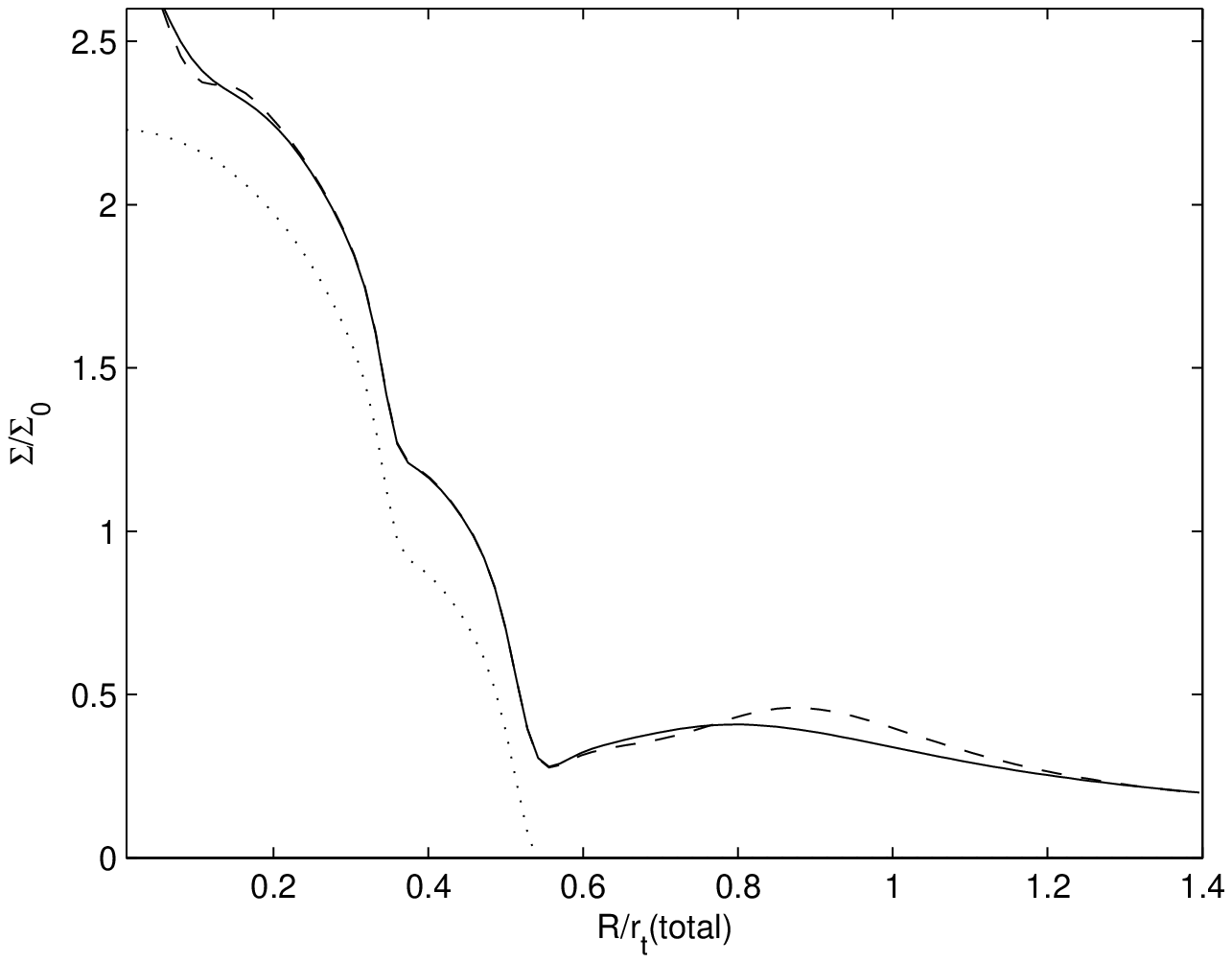} &
\includegraphics[width=0.5\columnwidth]{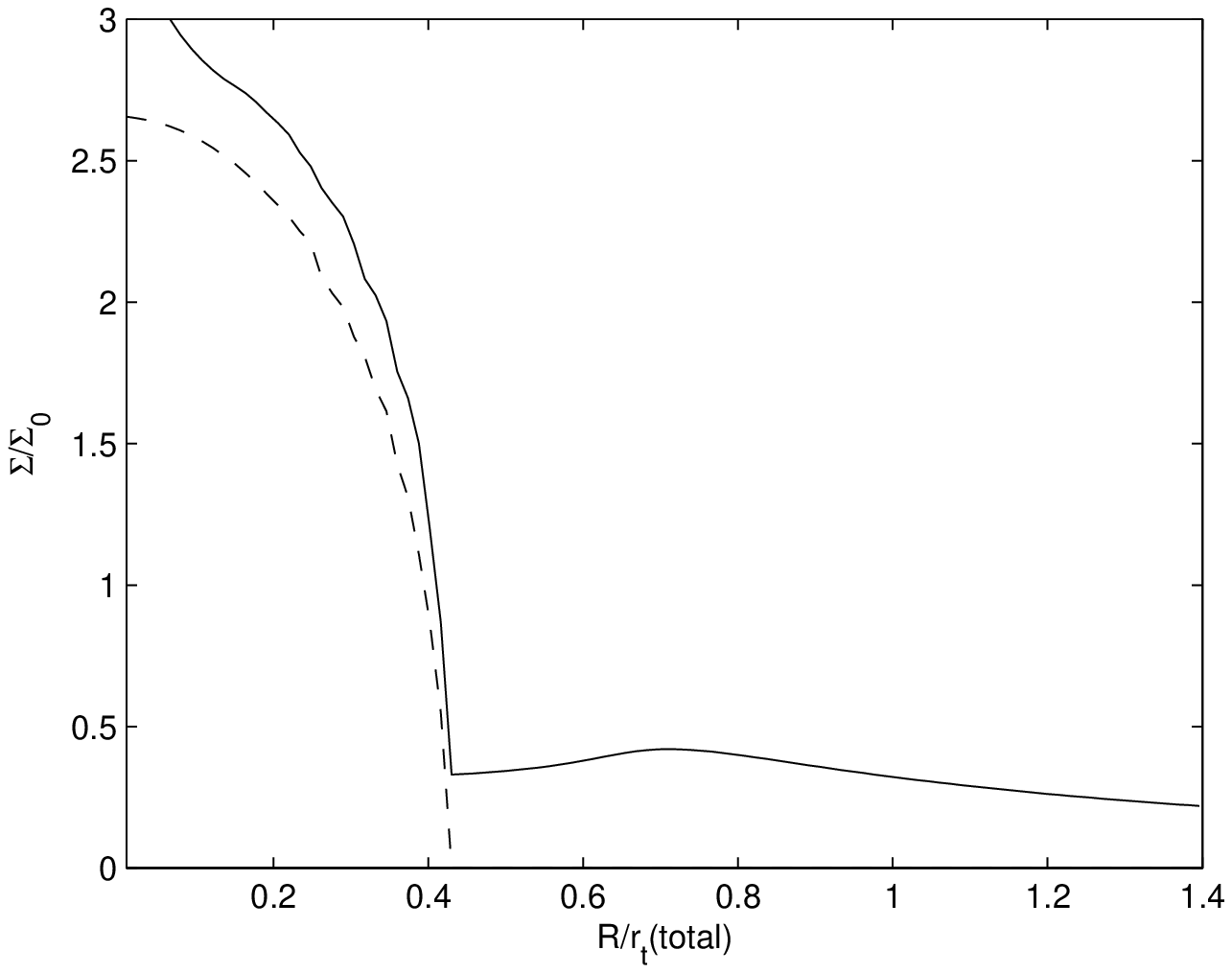}\\
\end{tabular}
\caption{The total projected Surface density in units of $\Sz$.
Upper left: for a single sphere of constant density with a radius
that is 0.6 the transition radius. Upper right: for two spheres of
constant density far apart from each other along the line of sight,
each has a radius that is 0.6 of its own transition radius. Lower
left: for two concentric spheres of constant densities of masses 1
and 0.3  and radii 0.53 and 0.35 of the total transition radius. All
these for two interpolating functions: $\bar\mu_2$ (solid) and
$\bar\mu_3$ (dashed). In each case the baryon contribution alone is
shown as the dotted line. Lower right: a dumbbell of two equal
spherical masses of constant density far apart along the line of
sight with $\mu_{10}$ (the source, baryon, contribution in dashed
line). For Cl 0024+17 the shear is $\kappa\approx 0.5\Sigma/\Sz$.}
 \label{fig6}
\end{figure}

As examples of a configurations that roughly reproduce the Jee et
al. parameters we show in Fig. \ref{fig6} the total projected mass
density of simple models for a few (favorable) forms of the
interpolating function. These are not generic but where
constructed to demonstrate that MOND can reproduce an observed
surface distribution like the one observed.

Concentrating on the dumbbell results in the upper Fig. \ref{fig6}
we see that $\S$ values comparable with those deduced by Jee et al.
can be achieved. Identifying the positions of the peak in the case
of this dumbbell in Fig. \ref{fig6} ($0.585r_t$, and $0.65r_t$ for
$\bar\mu_2$ and $\bar\mu_3$ respectively) with that observed
($400\kpc$) gives the total true (MOND) mass of the dumbbell as
$3.4\times 10^{14}\msun$, and $2.7\times 10^{14}\msun$,
respectively. According to our calculations, the integrated,
projected Newtonian dynamical mass (true plus phantom) within the
minimum in the above models is about 1.4  times the true source
mass; so, in absolute terms it comes out to $\sim 4.8\times
10^{14}\msun$, and $3.8\times 10^{14}\msun$, respectively, to be
compared with $4\times 10^{14}\msun$, which Jee et al. find within
the minimum. Mind you though that these are only indicative figures.
We also emphasize that the true (MOND) mass we require is still much
larger than the observed mass in gas and stars. From the $\beta$
model fit of Zhang et al. (2005), we estimate the gas mass within a
projected radius of 400 kpc to be $\sim 2.5\times 10^{13}\msun$; the
total mass of stars and gas together could be as large as $\sim
4\times 10^{13}$ M$_\odot$ (although it is difficult to estimate the
actual gas mass in this presumably double system). Therefore, as in
other clusters, MOND still requires a significant quantity of
undetected matter in the core (Sanders 2007). But the ``ring'' is
not made of this, of course.

Finally, we note that there are other mass configurations that can
achieve the same apparent surface density distribution; for
example, take the single sphere case in Fig. \ref{fig6} on the
background of a constant-surface density distribution along the
line of sight.

\section{Shells and rings in galaxies}

\begin{figure}
\begin{tabular}{rl}
\tabularnewline

\includegraphics[width=0.5\columnwidth]{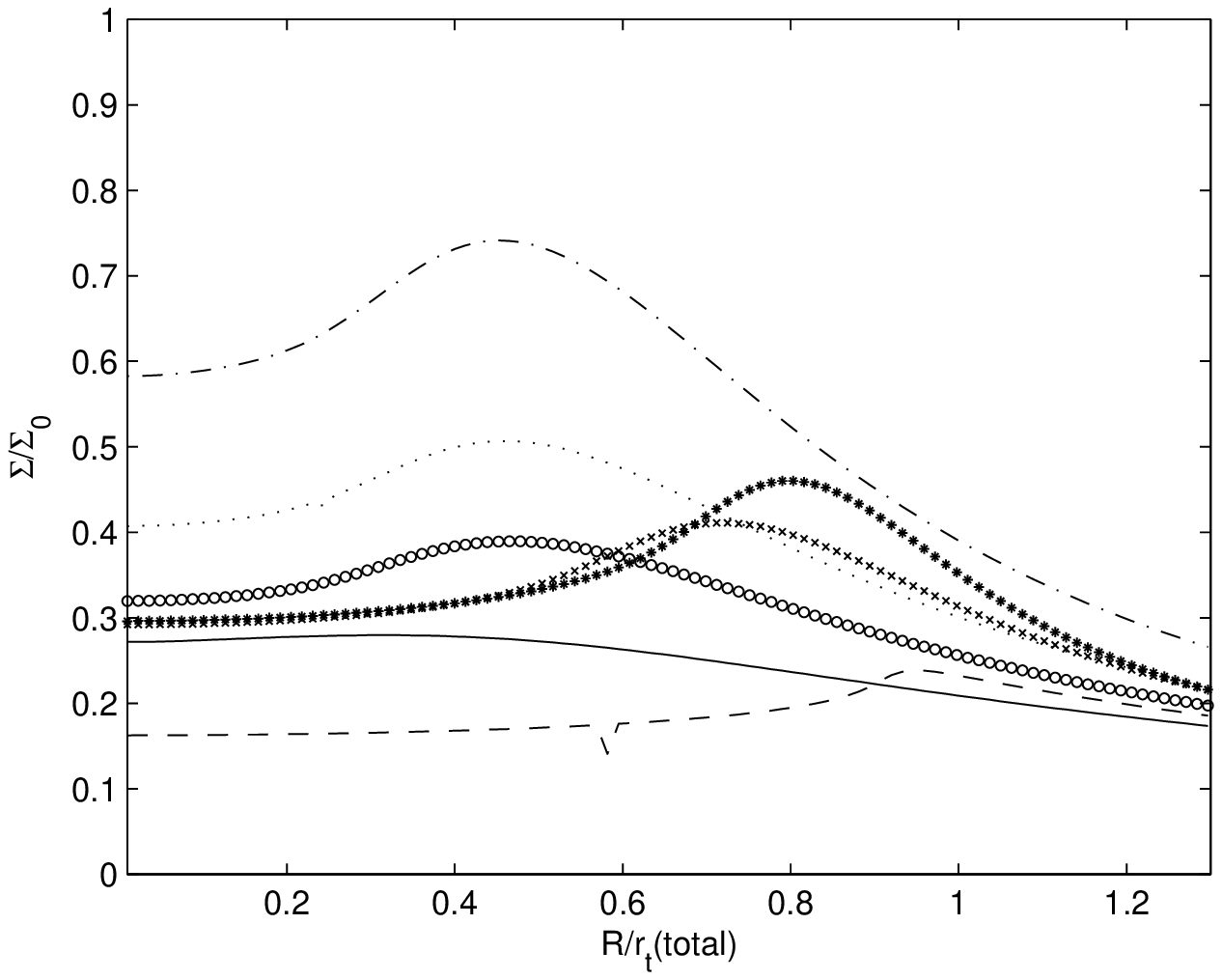} &
\includegraphics[width=0.5\columnwidth]{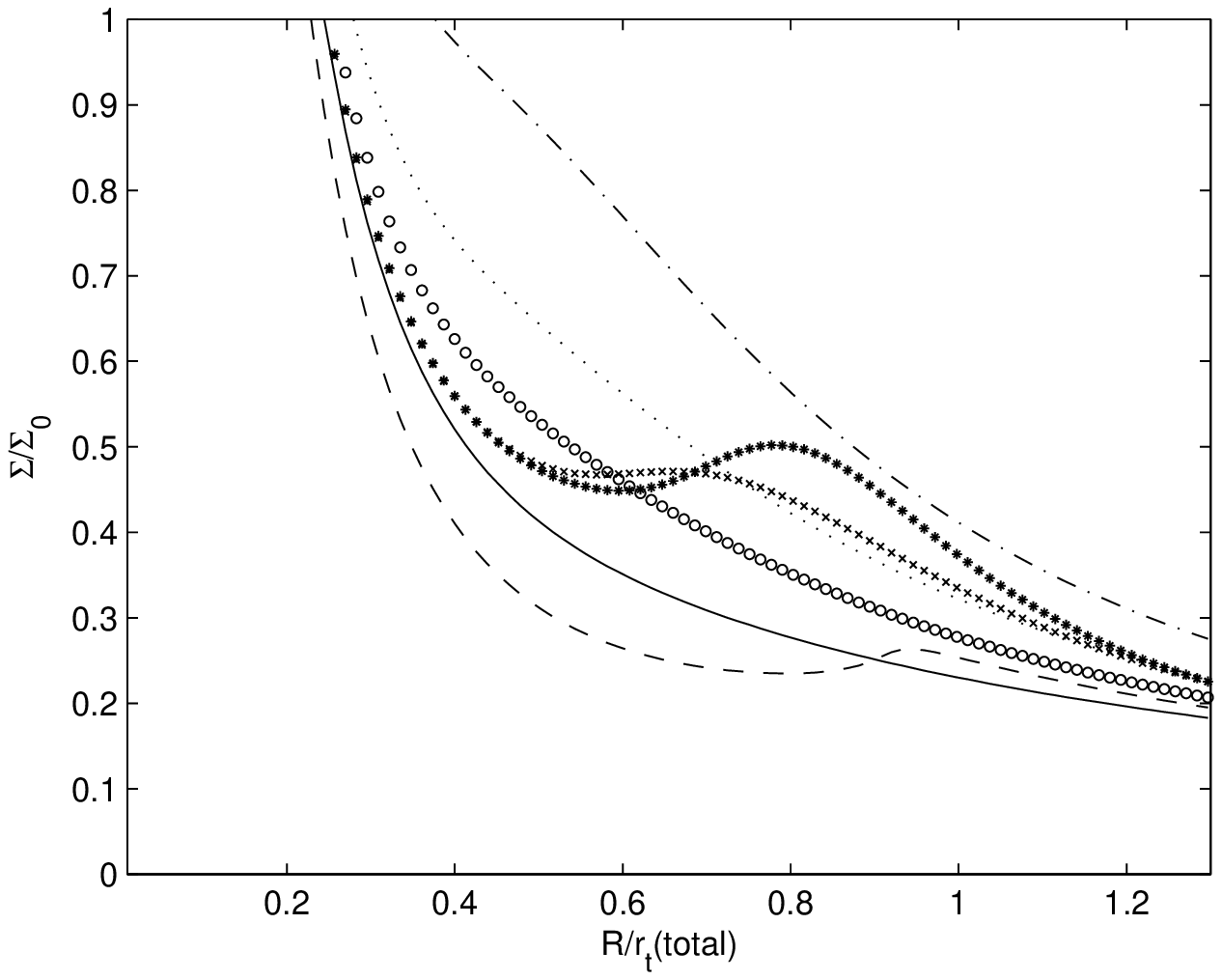}\\

\end{tabular}
\caption{The projected Surface density, in units of $\Sz$, for a de
Vaucouleurs model with $R_e=r_t/5.7 $, appropriate for NGC 3379. On
the left is shown the PDM alone, and or the right the total. The
interpolating functions used are: $\mu_2$ (solid), $\mu_{50}$
(dashed), $\tilde\mu_1$ (dotted), $\tilde\mu_2$ (dashed-dotted),
$\bar\mu_1$ (circles), $\bar\mu_2$ (crosses), $\bar\mu_3$ (stars).}
\label{fig7}
\end{figure}

The present discussion was prompted by the observation of a ring in
a galaxy cluster. But, of course,  the MOND orphan, transition
feature may appear in various ways also in  high surface brightness
(HSB) galaxies. A galaxy is dubbed HSB in the MOND sense if its
surface brightness is so high that the parameter $\xi\equiv
(MG/\ell^2\az)^{1/2}=r_t/\ell\gg 1$, where $\ell$ is the scale
length of the galaxy, say the half light radius for an elliptical
galaxy, or the exponential scale-length of a disc. In this case the
bulk of the baryonic mass is contained within the transition radius
and there is then a good chance for the appearance of a pronounced
feature. In fact, there is already evidence that in some HSB
galaxies the PDM distribution has a hole surrounding the center of
the galaxy. For example, this is the case for the Milky Way itself
(Bissantz, Englmaier \& Gerhard 2003). It has also been shown to
apply to some elliptical galaxies (Romanowsky et al. 2003, Milgrom
\& Sanders 2003). To demonstrate the expectations from one of the
galaxies in the study of Romanowsky et al. (2003), we show in Fig.
\ref{fig7} the projected surface density of the PDM alone and of the
total for a de Vaucouleurs sphere with $\xi=5.7$, appropriate for
NGC 3379 (Milgrom \& Sanders 2003). Had we had the means of
obtaining the dynamical surface density distributions of such
galaxies we would expect to see an analog of the ring in them (the
surface density is a semi-local quantity--density integrated only
along the line of sight--in which features appear more distinctly).
However, the presently applied methods of mapping the potential
field are not good at deducing the local surface densities. This is
certainly true of measuring the potential field via velocity
dispersion curves of test particles (as in Romanowsky et al. 2003).
The most accurate method to date for mapping the potential field of
disc galaxies is via rotation curve analysis. Can the transition
feature be seen directly on the rotation curve, which measures
volume integrated masses? Looking at Fig. \ref{fig2} we see that the
forms of interpolating function that are most conducive to the
visibility of a ring in the surface density distribution (e.g.
$\bar\mu_\alpha$ with $\alpha \ge 2$) could produce features in the
rotation curve that are not related to the underlying observable
mass distribution--i.e., orphan features--provided the galaxy is an
HSB one. Such a feature, which according to MOND can appear only in
HSB galaxies, and only at the location of the MOND transition, is to
be distinguished from the many known rotation-curve features whose
origin can be traced to features in the mass distribution. The
latter can appear in all galaxies and anywhere in a galaxy and,
unlike the orphan feature, they show up in the Newtonian rotation
curve (without DM). A famous case in point is the feature in NGC
1560 (Broeils 1992).

Of course, the transition from a Keplerian to a constant rotational
speed is itself a MOND transition feature; but, it is less
distinctive than a marked dip, and much less discriminative among
the various forms of the interpolating function.

Until recently, almost all disc galaxies with reported rotation
curves and MOND analysis had $\xi\le 1$; for these the form of the
interpolating function makes very little difference in the
predicted rotation curve; in particular, there is no prediction of
an appreciable orphan feature on the rotation curve. This lack of
data and analysis has been remedied by Sanders \& Noordermeer
(2007). They presented MOND analysis of a sample of HSBs, using
$\mu_1$ for their MOND fits; this, as we saw, does not produce a
distinct feature even under the most favorable conditions [Figs.
\ref{fig2}-\ref{fig3}, and the discussion below eq.(\ref{x})].

On the other hand there where indeed some features on the observed
rotation curves that where not reproduced by the MOND curves in the
analysis of Sanders and Noordermeer (2007). (These are also not
explained by cold dark matter (CDM) fits, such as NFW profiles; see
Noordermeer 2006.) We have now analyzed some of these galaxies with
choices of $\mu$ from our new arsenal. Interestingly, some these
reproduce the previously unexplained orphan features.

\begin{figure}
\begin{tabular}{rl}
\tabularnewline
\includegraphics[width=1.0\columnwidth]{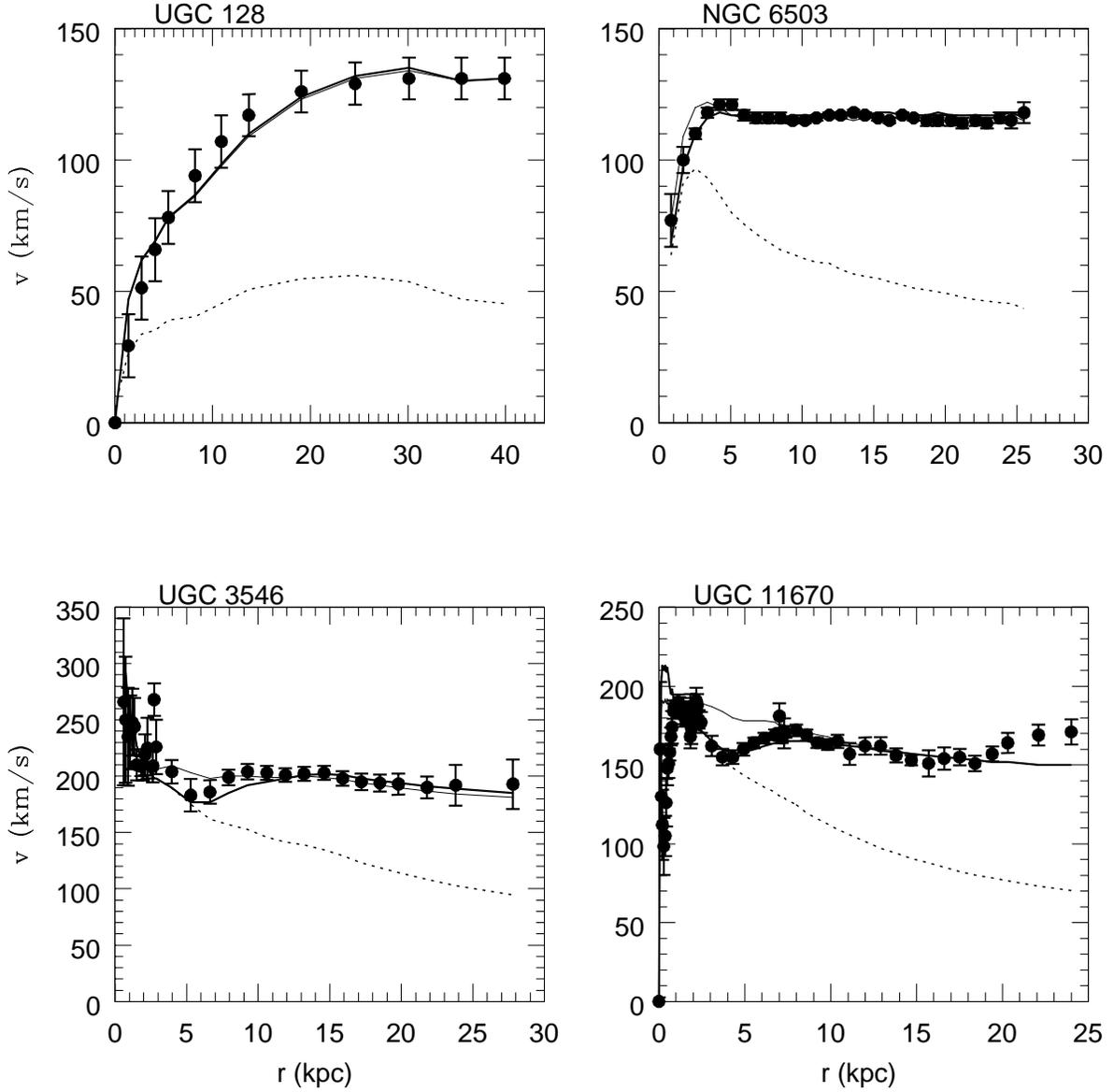}  \\
\end{tabular}
\caption{MOND rotation curves for four galaxies.  UGC 128 is a low
surface brightness galaxy (Sanders 1996); NGC 6503 is a typical
spiral galaxy with average surface brightness (Begeman et al. 1991);
UGC 3546 and UGC 11670 are two high surface brightness galaxies from
the sample of Noordermeer (Sanders \& Noordermeer 2007).  The points
with error bars are the observations. Also shown are the calculated
rotation curves: MOND for $\bar\mu_2$ (solid thick), and MOND for
$\mu_1$ (solid thin). The best fit $M/L$ values in solar units are:
1.0 for UGC 128 (disc only, both fits); 0.9 for NGC 6503 (disc only,
both fits). UGC 3546: 2.5 (disc) and 7.0 (bulge) for $\bar\mu_2$,
2.5 (disc) and 5.9 (bulge)  for $\mu_1$. UGC 11670: 2.5 (disc) and
4.5 (bulge) for $\bar\mu_2$, 3.0 (disc) and 3.5 (bulge) for $\mu_1$.
Newtonian curves (dotted) are given for the $M/L$ values from the
$\bar\mu_2$ fits.}
 \label{fig8}
\end{figure}

Figure \ref{fig8} shows the observed rotation curves of four spiral
galaxies: UGC 128, a low surface brightness (LSB) galaxy with
$\xi<1$ (Sanders 1996); NGC 6503, a ``normal'' disc galaxy with $\xi
\approx 1$ (Begeman et al. 1991); and two HSB galaxies from the
sample of Sanders \& Noordermeer (2007) with $\xi>1$.  We show the
curves predicted by MOND from the observed mass distribution using
both $\mu_1$ (reproducing the fits of Noordermeer and Sanders) and
$\bar\mu_2$. We see that for the first two galaxies (with $\xi\le
1$) there is very little difference between the two predictions; and
there are no orphan features to be found. However, for the two HSB
galaxies, there is a clear indication of an orphan feature in the
observed curves at the point near the transition from the Newtonian
regime to the MOND regime. Moreover, this feature is reproduced by
$\bar\mu_2$ but not by $\mu_1$. However, E. Noordermeer (2007,
private communication) warns that these galaxies are barred and that
the dips in the rotation curves may be due to non-circular motions,
and not to true features of the potential field.
\par
\begin{figure}
\begin{tabular}{rl}
\tabularnewline
\includegraphics[width=1.0\columnwidth]{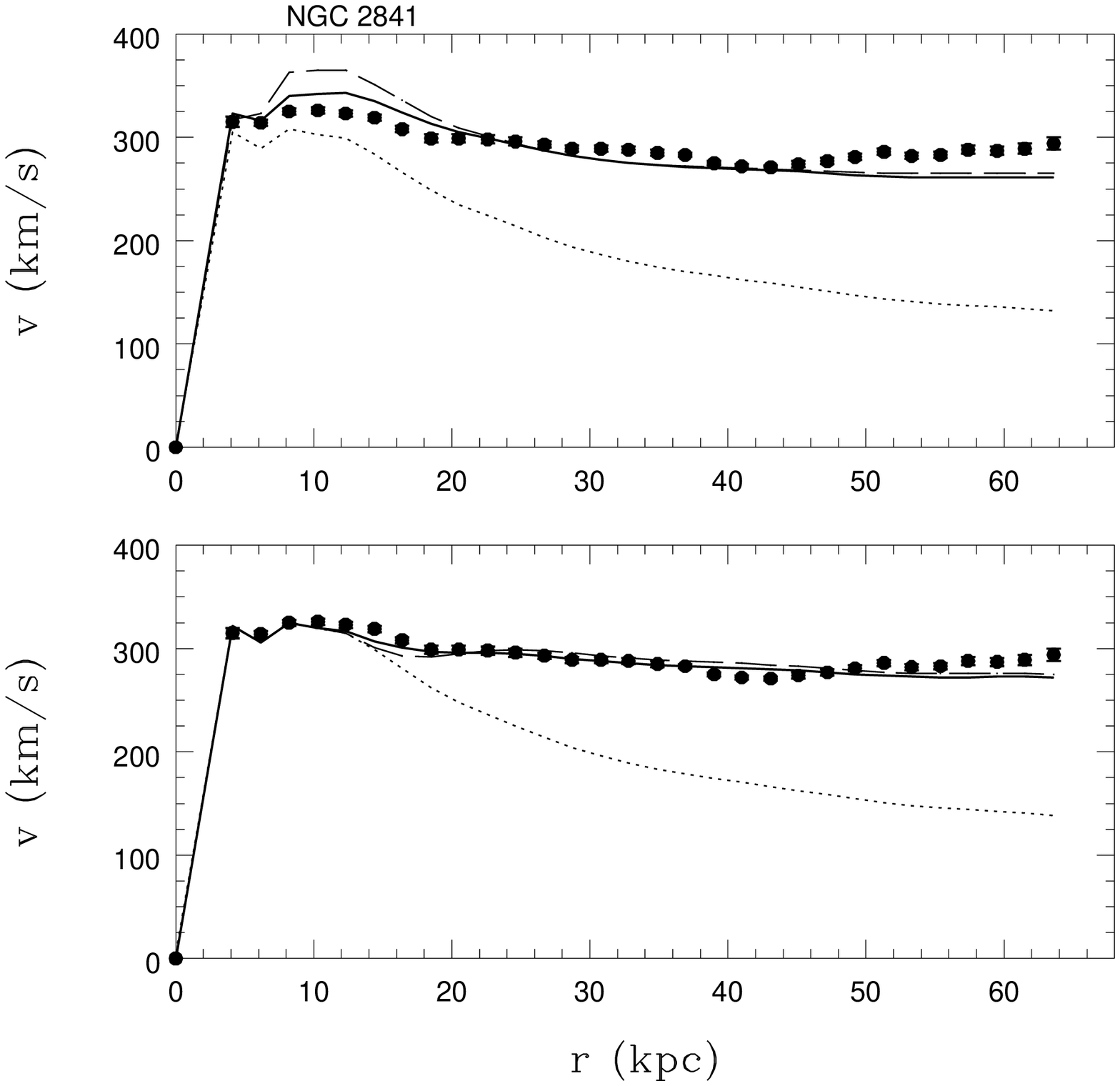}  \\
\end{tabular}
\caption{The MOND rotation curves of NGC 2841 for different choices
of the interpolating function compared with the data (points). The
best fit $M/L$ values in solar units are given below in parentheses
for the disc and bulge respectively. Upper panel: $\mu_2$ in dashed
line (8.1, 1.0); $\mu_1$ in solid line (4.9, 2.5); Newtonian curve
with $M/L$ values for $\mu_1$ (dotted). Lower panel: $\bar\mu_2$ in
dashed line (5.4, 2.8); $\bar\mu_{1.5}$ in solid line (5.4, 2.8);
Newtonian curve (dotted).}
 \label{fig9}
\end{figure}

While the MOND rotation curves of LSB galaxies are quite insensitive
to the form of the interpolating function, those of HSB galaxies are
rather sensitive. This raises the possibility that with the new
forms of $\mu$ studied here we may solve a long standing problem
concerning NGC 2841. This is an HSB galaxy that has posed a puzzle
in the context of MOND. Begeman Broeils \& Sanders (1991), and more
recently Bottema \& al. (2002), found that a very good MOND fit for
the rotation curve of this galaxy is obtained if the galaxy is at a
distance somewhat larger than 20\mpc. Putting it at its Hubble
distance of $9.46\mpc$ (as in the former analysis), or even at the
more recently determined Cepheid distance of $14.1\pm1.5\mpc$ (as in
the latter analysis) gave an inferior fit. Both analyses used
$\mu_2$. We now find that with some of our new interpolating
functions, which are also expedient for ring formation, the MOND fit
is very good for the Cepheid distance of NGC 2841. We show in Fig.
\ref{fig9} the fit with $\mu_2$, reproducing the old conundrum, and
we see that using $\mu_1$ alleviates the discrepancy somewhat but
not completely. The forms $\bar\mu_2$ and $\bar\mu_{1.5}$ do a much
better job. This may then be the correct solution to the puzzle: not
to put NGC 2841 at a larger distance but to use a more appropriate
form of $\mu$ (a combination of the two is also possible, of
course). It should be noted, however, that the larger distance of
$23\mpc$ is indicated both by the Tully-Fisher (TF) relation and by
a type Ia supernova in NGC 2841; so the issue remains moot.
\par
To summarize, we may have evidence from rotation curves that the
form of the interpolating function conducive to the appearance of a
ring of phantom matter, may also be that which is appropriate for
the matching of rotation curves in HSB (and LSB) discs. We can also
tentatively identify the previously unexplained dips in the observed
rotation curves of UGC 3546 and UGC 11670 as the orphan features
associated with the MOND transition.

\par
Famaey \& Binney (2005) found in their MOND analysis of the Milky
Way rotation curve, that it is best fitted with a $\mu$ form that
starts as $\mu_1(x)$ at small $x$, then increase faster and become
similar to $\mu_2(x)$ at larger $x$. Our $\bar\mu_2,~\bar\mu_3$ also
start as $\mu_1$ at small x and then go quickly to 1 (but rather
more quickly than in the form used by Famaey and Binney).

\par
The time is becoming ripe for a comprehensive MOND analysis of
rotation curves, and other data, to better constrain the form of the
interpolating function. The inclusion of the consideration of the
orphan, MOND-transition feature adds a new dimension to this sort of
analysis.

\section{Summary and Discussion}
We have shown that, under certain conditions, the PDM distribution
in MOND exhibits a maximum along some surface whose scale is largely
determined by the MOND transition radius of the underlying
(baryonic) mass. This appears without there being a corresponding
feature in the underlying source distribution. In axisymmetric
configurations aligned with the line of sight this can appear as a
ring of DM at that radius, which is unrelated to a length scale
characterizing the source. This is expected to hold in any theory of
modified dynamics (not only MOND) at the radius that marks the
transition from the Newtonian regime to the modified regime.
For example, in theories where the transition to modified gravity
occurs at a fixed length scale, the feature would appear at
that length scale regardless of the mass of the object.

Thus, the statement of Jee et al. (2007) to the effect that the
appearance of such an orphan feature is a direct proof of DM, and
that it disagrees with modified dynamics interpretation, is
incorrect. In fact, the observed parameters of the observed ring in
the galaxy cluster Cl 0024+17 (radius, surface density, contrast,
and the mass involved) can be naturally reproduced with this pure
MOND phenomenon. The Jee et al. ring may thus turn out to constitute
a direct evidence for MOND in action.

For a mass that is well contained within its transition radius, the
ring appears quite generically and for a wide range of MOND
interpolating functions. The exact properties of the ring depend,
however, sensitively on the choice of interpolating function; if
enough rings of this type are detected important constraints on that
function may emerge. Because the surface density of the ring cannot,
generically, much exceed $\Sz$ it is easily overwhelmed by the
source distribution itself, and so is not expected to appear very
commonly in the {it total} mass distribution of realistic
configurations (unless we can confidently subtract the underlying
source distribution). The properties of the feature also depend
strongly on the distribution of the source mass along the line of
sight, demonstrating clearly the gross inapplicability of the
thin-lens approximation in MOND.

We have not considered non-axisymmetric configurations, but it is
expected that in such cases the feature will appear distorted in
shape, perhaps broken or irregular.

Furthermore, we found preliminary evidence of a conspicuous orphan
feature (a dip) on the rotation curves of some HSB galaxies, that
can be identified with the transition from the Newtonian regime to
the MOND regime-- a feature predicted by MOND with just the forms of
the interpolating function that are most conducive in our sample to
the appearance of a ring in the PDM surface density distribution.
This calls for a reconsideration of the rotation curves of spiral
galaxies using these alternative forms of the interpolating
function, and possibly others, as well as a systematic study of the
implied PDM density distribution in individual galaxies.

 Finally, it
should be recalled that we have used the algebraic relation, which
allowed us to treat and compare many cases, but which for
non-spherical systems is only approximate.  In a proper treatment of
gravitational lensing in the context of a theory inspired by TeVeS,
one should solve the non-linear Bekenstein-Milgrom field equation
for a given mass distribution.

\acknowledgements
This research was supported, in part, by a
center of excellence grant from the Israel Science Foundation.

\end{document}